\documentclass[10pt,twocolumn,twoside]{IEEEtran}

\usepackage{amsmath} 
\usepackage{amssymb} 

\expandafter\def\expandafter\normalsize\expandafter{%
    \normalsize
    \setlength\abovedisplayskip{3pt}
    \setlength\belowdisplayskip{3pt}
    \setlength\abovedisplayshortskip{3pt}
    \setlength\belowdisplayshortskip{3pt}
}

\usepackage{amsthm}
\usepackage{graphicx}
\usepackage[space]{grffile}

\usepackage{cite}

\usepackage{wrapfig}

\DeclareMathOperator*{\argmin}{arg\,min}

\title{\LARGE \bf
Optimal Triggering of Networked Control Systems
}

\author{Ali Heydari$^1$
\thanks{$^{1}$Assistant Professor of Mechanical Engineering, South Dakota School of Mines and Technology, Rapid City, SD 57701, phone: 605-394-2200, email: ali.heydari@sdsmt.edu.}}

\newtheorem{Thm}{Theorem} 
\newtheorem{Lem}{Lemma} 
 
\newtheorem{Def}{Definition} 
\newtheorem{Assumption}{Assumption} 

\begin{document}

\maketitle
\pagenumbering{arabic}
\pagestyle{plain}
\thispagestyle{plain}

\begin{abstract}
The problem of resource allocation of nonlinear networked control systems is investigated, where, unlike the well discussed case of triggering for stability, the objective is \textit{optimal} triggering. An approximate dynamic programming approach is developed for solving problems with fixed final times initially and then it is extended to infinite horizon problems. Different cases including Zero-Order-Hold, Generalized Zero-Order-Hold, and stochastic networks are investigated. Afterwards, the developments are extended to the case of problems with unknown dynamics and a model-free scheme is presented for \textit{learning} the (approximate) optimal solution. After detailed analyses of convergence, optimality, and stability of the results, the performance of the method is demonstrated through different numerical examples.
\end{abstract}

\section{Introduction}
\begin{wrapfigure}{r}{0.28\textwidth}
  \vspace{-7pt}
  \begin{center}
		\includegraphics[width=0.55\columnwidth]{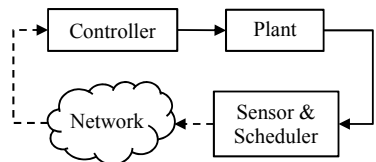}
  \end{center}
  \vspace{-10pt}
		\caption{Schematic of an NCS with the sensor connected to the controller through a network. Dash lines transmit data only when the network is scheduled/triggered.} 
		\label{Fig1_NCS}
  \vspace{-8pt}
\end{wrapfigure}

Unlike conventional control systems, the control loop is closed through a \textit{communication network} with a \textit{limited bandwidth} \cite{Hristu_LimitedCommunication} in a \textit{Networked Control System} (NCS), \cite{Halevi_NSC_1988_DelayMainly, Walsh_NSC_ContSystMagazine, Yook_Event_Driven_StateTHreshold, Ishii_SwitchBox, Antsaklis_PeriodicTriggering_ModelBased, Yang_NCS_ShortSurvey, Antsaklis_NCS_SpecialIssue, Gupta_NCS_Overview} as shown in Fig. \ref{Fig1_NCS}.
While a traditional feedback controller requires constant (or periodic) access to the sensor, and hence, to the network for receiving state measurements, several different tasks are \textit{competing} for transmitting their data through the same network in an NCS. Therefore, designing schemes which decrease the communication load of the network while maintaining the desired performance is beneficial, \cite{Walsh_NSC_ContSystMagazine, Antsaklis_MB_NCS_EVentTriggering}.

As a real-world example, smart power grids with small-scale electricity generators may be mentioned, in which not only the sensors located at points of common coupling are spatially distributed throughout the network, but also, the generators are distributed, \cite{NCS_SmartGrid}. This leads to a flow of electricity and information in the power network. 
As another example, different control systems of an airplane, e.g., flight control, engine control, etc., can be considered. Replacing the wire harnesses of the sensors, which are spatially distributed throughout the plane, with a unified, potentially wireless, communication medium leads to a dramatic decrease in the weight and hence, the operation cost. Simultaneously, it will facilitate the maintenance and improve the monitoring and fault detection capabilities, \cite{WirelessSensorNetwork_Airplane}. In a simple and small airplane (Cessna 310R) this change was reported to increase the range by 10\%, \cite{WirelessSensorNetwork_Airplane}. 

Considering the literature, the main developed approaches are 
a) decreasing the need for continuous state measurement through designing periodic \cite{Antsaklis_PeriodicTriggering_ModelBased, Ishii_SwitchBox	, Nesic_QCS_NCS} or aperiodic event triggering schemes \cite{Astrom_ProposingEventTriggering, Yook_Event_Driven_StateTHreshold, Walsh_TOD, Tabuada_EventTriggering, Astorm_GZOH, Rabi_OptimalControl_NCS, Heemels_IJC_ZOH, Lunze_PerfectModel_GZOH, Wang_ISS, Marchand_ZOH, Cassandras_CCC, Antsaklis_MB_NCS_EVentTriggering, Sahoo_Jagannathan_ACC13} where the network is utilized only when an event is triggered, 
b) decreasing the size of the data packets for reducing the network load \cite{Tatikonda_EncoderDecoder, Premaratne_EventTriggering_SpeechCoding}, and 
c) designing controllers which `deal' with the busy network and its consequences including induced delays \cite{Halevi_NSC_1988_DelayMainly, Goktas_Delay_RobustControl, Nilsson_LQG_forDelays_TimeStamping, 
Wu_NCS_Delay, Yue_RobustCOntrol_For_Delayand_Droppout, Kim_MaximumAllowableDelayBound, Yi_BackProp_ForDelayApprox, Du_AMC_AdaptiveCritic_StochasticSystem, Xu_Jagannathan_Automatica,  Repele_Delay_SmithPredictor_DelayCalculation} or quantization errors \cite{Delchamps_Quantization, Brockett_Quantization} which exist anyway in digital networks \cite{Nesic_QCS_NCS}. While the idea of compressing data packets was observed not to be as effective as the idea of transmitting less frequently \cite{Antsaklis_MB_NCS_EVentTriggering}, the first and last approaches have been very popular, especially if the two approaches can be combined to both transmit less frequently and account for the losses and delays, \cite{Antsaklis_MB_NCS_EVentTriggering}.

Event triggering, as opposed to periodic sampling, has been typically conducted through monitoring the error between the current state of the system and the state information expected to be available on the other side of the network and triggering the system (or scheduling the network) when the error exceeds a certain limit. The state information available on the other side of the network could be the last state measurement transmitted through the network as in Zero-Order-Hold (ZOH) based methods \cite{Tabuada_EventTriggering, Heemels_IJC_ZOH, Wang_ISS, Marchand_ZOH} or an estimation of the current system state made based on the last transmitted state measurement, as in Generalized ZOH (GZOH) or Model Based NCS schemes \cite{Antsaklis_PeriodicTriggering_ModelBased, Antsaklis_PeriodicTriggering_With_TimeVaryingPeriod, Astorm_GZOH, Lunze_PerfectModel_GZOH, Antsaklis_MB_NCS_EVentTriggering}. In any case, the controller is designed a priori with the assumption of constant access to the sensor, i.e., without considering the event triggering nature, and the triggering/scheduling policy (sometimes called the \textit{event function}) is designed such that the system is stabilized, in the cited papers.

Reviewing the available literature, the area of triggering optimally as opposed to triggering for stability is rarely investigated due to its difficulties, despite its natural advantages. The published studies in this area, to the best of the author's knowledge, includes results in \cite{Xu_CDC2004_OptimalCommunicationPolicy, Farokhi_TAC, Pappas_TAC_OptimalPOwerManagement} for linear systems. This study is aimed at this pursuit. 

The contribution of this study is extending the applications of Approximate Dynamic Programming (ADP) \cite{Werbos_ActionDependent_Critic, Bertsekas_NDP} and Reinforcement Learning (RL) \cite{Watkins, Sutton} to (near) optimal triggering of NCSs. Initially the case of finite-horizon cost functions with the simplifying assumption of applying no control, when no state feedback is received by the controller is investigated. This assumption simplifies the problem to a switching problem, hence, the method developed in \cite{Heydari_Franklin} is directly applicable. Once the idea behind the solution is clarified, the more advanced cases of ZOH and GZOH are covered and the outline of the process for extending the results to stochastic networks with random delays and losses is given. Afterwards, the schemes are extended to the case of infinite-horizon problems and both model-based and model-free schemes are developed. These methods are supported by rigorous analyses on convergence, optimality, and stability. 

The developed methods lead to low online computational load. Also, they are valid for different initial conditions of the system. Moreover, the schemes are \textit{scalable} and \textit{decentralized}. That is, when implemented on systems with several sensor-controller-plant sets sharing the same network, the scheduling is based solely on the states of the respective plant for each set, so, it can be conducted \textit{in parallel} for every set. 

As for the approach of this study, ADP has already been investigated for control of NCSs in \cite{Xu_Jagannathan_Automatica,HaiboHe_NCS,Pappas_TAC_OptimalPOwerManagement}. 
However, ADP was used for finding the control law in \cite{Xu_Jagannathan_Automatica} and \cite{HaiboHe_NCS} under some ADP-\emph{independent} triggering policies. Also, the use of ADP in adjusting the transmission power (which changes continuously) was presented in \cite{Pappas_TAC_OptimalPOwerManagement}. In here, though, ADP will be used for optimal design of the event function. 
Finally, an ADP based approach to optimal intermittent feedback was presented in \cite{Ferrari_EventTriggering}. The approach involves an online gradient descent search for finding the next optimal time for triggering, based on the value function representing the cost. While the authors presented their initial and interesting developments to the problem in that paper, no further work appeared from them in deeper analysis of the method, to the best of this author's knowledge.

The rest of this paper is organized as follows. The problem is formulated in the next section, followed by the idea for solving it. Afterward, the extensions of the work to ZOH, GZOH, stochastic networks, infinite-horizon problems, and problems with modeling uncertainty are presented in the subsequent sections. Numerical examples are presented in Section \ref{Simulations}, followed by some concluding remarks.

\section{Problem Formulation}  \label{Problem_Formulation}
The problem of \textit{scheduling} an NCS can be presented as follows. Let the discrete-time dynamics of the plant in Fig. \ref{Fig1_NCS} be given by
\begin{equation}
x_{k+1} = f(x_k,u_k), \forall k \in \mathbb{K}:=\{ 0,1,...,N-1 \},\label{Dynamics}
\end{equation}
where $f:\mathbb{R}^n \times \mathbb{R}^m \to \mathbb{R}^n$ is a continuous function versus its both inputs, i.e., the state and control vectors, $x$ and $u$, respectively, with $f(0,0)=0$. Sub-index $k$ represents the discrete time index, $N$ denotes the fixed final time, and positive integers $n$ and $m$ denote the dimensions of the continuous state and control spaces.

Let the piecewise constant function $v_k:\mathbb{R}^n \to \{0,1\}$ denote the \textit{scheduling/triggering policy}, where $v_k(x)=1$ means the event is triggered or equivalently, the network is \textit{scheduled} to be utilized for transmitting state information through the dash lines in Fig. \ref{Fig1_NCS}, and $v_k(x)=0$ means the network is not scheduled (event is not triggered). Assume cost function
\begin{equation}
J= \psi(x_N) + \sum_{k=0}^{N-1} U(x_k,v_k), \label{CostFunction_FinHor}
\end{equation}
where continuous positive semi-definite function $\psi:\mathbb{R}^n \to \mathbb{R}^+$ assigns a \textit{final cost} to the last state vector and positive semi-definite $U:\mathbb{R}^n \times \{0,1\} \to \mathbb{R}^+$, continuous versus the first argument, assigns a \textit{running cost} to the states during the horizon and to each network scheduling. For example, one may select $U(x,v) = Q(x) + c v$ for a some $Q:\mathbb{R}^n \to \mathbb{R}^+$ and a positive constant $c$, where the latter assigns a weight to the cost of usage of the resource, i.e., the network. The set of non-negative reals is denoted with $\mathbb{R}^+$. 

Moreover, let function $h: \mathbb{R}^n \to \mathbb{R}^m$ denote a known continuous feedback control policy, i.e., $u_k = h(x_k), \forall k$, such that it asymptotically stabilizes the system if the network is always scheduled. 
The problem is designing a feedback triggering policy $v_k(.), \forall k \in \mathbb{K}$, such that cost function (\ref{CostFunction_FinHor}) is minimized, subject to plant (\ref{Dynamics}) under control policy $h(.)$. 

\section{Basic Idea for the Solution}  \label{BasicProblem}

In order to demonstrate the basic idea behind the solution presented in this study, let the control be zero when the network is not scheduled, i.e., $v_k = 0 \Rightarrow u_k=0$, or simply $u_k=v_k h(x_k)$, called No Feedback, No Control (NFNC), in this study. Also, the dynamics of the system are assumed to be known and the sensor node is assumed to have a copy of control policy $h(.)$. Moreover, The network is assumed to be delay-free and loss-free.

Motivated by switching systems \cite{Heydari_Franklin}, the operation of the system can be modeled by \textit{switching} between two different \textit{modes}; $F_0(.)$ and $F_1(.)$, respectively, for the case of $v_k=0$ and $v_k=1$,
\begin{equation}
\begin{split}
&x_{k+1}=F_0(x_k) := f\big(x_k,0\big) ,\\
&x_{k+1}= F_1(x_k) := f\big(x_k,h(x_k)\big). \label{Modes_SimpleIdea}
\end{split}
\end{equation}
The idea for solving this switching problem is using the Bellman equation \cite{Kirk}. Let the \textit{optimal value function}, sometimes called \textit{optimal cost-to-go}, at current state $x_k$ and time $k$ which leads to \textit{time-to-go} $\tau := (N-k)$, be denoted with $V^*_{\tau}(x_k)$ for some $V^*_{\tau}:\mathbb{R}^n \to \mathbb{R}^{+}, \forall \tau \in \mathbb{K} \cup {N}$. By definition,  optimal value function $V^*_{\tau}(x_k)$ represents the incurred cost from the current time to the final time if \textit{optimal decisions} are made throughout the remaining $\tau$ time steps. Considering Eq. (\ref{CostFunction_FinHor}) the value function satisfies recursion $V^{*}_{\tau}(x_k)= U(x_k,v_k^*) + V^{*}_{\tau-1}\big(F_{v_k^*}(x_k) \big)$, where $v_k^*  \in \{0,1\}$ denotes the optimal decision at time $k$. Based on the Bellman equation, one has
\begin{equation}
V^{*}_0(x)=\psi(x), \forall x, \label{Bellman_FinHor_eq1}
\end{equation}
\begin{equation}
V^{*}_{\tau}(x)= \min_{ v\in\{0,1\} } \Big( U(x,v) + V^{*}_{\tau-1}\big(F_v(x) \big) \Big), \forall x, \forall \tau \neq 0. \label{Bellman_FinHor_eq2}
\end{equation}
If the value function is available, then, the \textit{{optimal triggering policy}} at time $k$ (which leads to the time-to-go $\tau=N-k$) with current state $x$ is simply given in a feedback form by
\begin{equation}
v_k^*(x)= \argmin_{ v\in\{0,1\} } \Big( U(x,v) + V^{*}_{N-(k+1)}\big(F_v(x) \big) \Big). \label{Scheduler_eq1}
\end{equation}
Interestingly, Eqs. (\ref{Bellman_FinHor_eq1}) and (\ref{Bellman_FinHor_eq2}) can be used for approximating the desired value function. Utilizing any parametric function approximator, e.g., Neural Networks (NN), with \textit{time-dependent parameters/weights} for approximating the time-dependent value function, the approximation can be done in a backward fashion, from $k=N$ to $k=0$, or equivalently from $\tau=0$ to $\tau=N$. 
Eq. (\ref{Bellman_FinHor_eq1}) can be used for finding the parameters of $V_0^*(.)$ approximator, e.g., using least squares as detailed in \cite{Heydari_Franklin}, within some selected compact domain of interest $\Omega$, i.e., $\forall x\in\Omega$. Afterwards, having $V_0^*(.)$, Eq. (\ref{Bellman_FinHor_eq2}) can be used for approximating $V_1^*(.)$. Repeating this process all the functions can be approximated in the offline phase. Afterwards, the optimal triggering can be conducted in realtime using Eq. (\ref{Scheduler_eq1}). This calculation is as simple as evaluating two scalar valued functions and selecting the $v$ which corresponds to the smaller number.

This method depends on the approximation accuracy of the function approximator. NN, as parametric function approximators, are known to provide \textit{uniform approximation} with any desired accuracy providing the function subject to approximation is a \textit{continuous} function, \cite{Weierstrass_Theorem, Hornik_NN_Continuity}. Considering Eq. (\ref{Bellman_FinHor_eq1}), the continuity of the function subject to approximation follows from the continuity of $\psi(.)$. For Eq. (\ref{Bellman_FinHor_eq2}), however, due to the switching between $v=0$ and $v=1$, the continuity versus $x$ is not obvious. This continuity is established in an earlier work on optimal switching, in \cite{Heydari_Franklin}.

\section{Extension to Zero-Order-Hold} \label{ZOH_FinHor}

In Section \ref{BasicProblem} the simplified problem of having $u_k=0$ when $v_k=0$, NFNC, was investigated. In most of the case, this is not as good as at least applying the previously calculated control when no new feedback information is received, called \textit{Zero-Order-Hold} (ZOH), \cite{Tabuada_EventTriggering, Heemels_IJC_ZOH, Wang_ISS, Marchand_ZOH}. Let the \textit{last} state measurement received by the controller \textit{as of time} $k$, be denoted with $d_k$. In ZOH one has $u_k=h(d_k)$ when $v_k = 0$, and has $u_k=h(x_k)$ when $v_k = 1$.
Also, let the finite-horizon cost-function (\ref{CostFunction_FinHor}) be assumed. The objective is designing the scheduling policy $v_k$ such that cost function (\ref{CostFunction_FinHor}) is minimized, under the ZOH policy of the controller node. 

Once the controller has a \textit{memory} to store the last received state measurement $d_k$, it can be seen that the value function will be dependent on the stored $d_k$ as well as on time and $x_k$. To clarify this point, two cases may be considered, where, in both cases the current states and times are the same, but, the last received state measurements, i.e., $d_k$s are different. The cost-to-go can then be different, because, if the schedulers decide not to schedule the network in both cases, then, the controls that the controllers will apply on their respective plants will be different due to the different $d_k$s, and this can lead to different $x_{k+1}$s. Another way of looking at this dependency, is considering $d_k$ also as a part of the \textit{state} of the system. The \textit{overall state vector}, denoted with $y_k\in\mathbb{R}^{2n}$, which is supposed to uniquely identify and characterize the current status of the system includes both the current physical state, $x_k$, and the last transmitted state measurement, $d_k$, i.e., $y_k:=[x_k^T, d_k^T]^T$. Considering this argument, the value function of the NCS with ZOH may be denoted with $V^*_{\tau}(y_k)$ for some $V^*_{\tau}:\mathbb{R}^{2n} \to \mathbb{R}^{+}, \forall \tau \in \mathbb{K} \cup {N}$.

Let the dynamics of $y_k$ under the two events of the network not being scheduled and being scheduled, be denoted with modes $F_0(.)$ and $F_1(.)$, respectively. Then,
\begin{equation}
\begin{split}
y_{k+1}=F_0(y_k) := \left[ 
		\begin{array}{c}
			f\big(x_k,h(d_k)\big) \\
			d_k
		\end{array}
		\right], \\
y_{k+1}= F_1(y_k) := \left[ 
		\begin{array}{c}
			f\big(x_k,h(x_k)\big) \\
			x_k
		\end{array}
		\right].		
		\label{Dynamics_ZOH_y_k}
\end{split}
\end{equation}
For example, as seen in $F_1(.)$, when the network is scheduled one has $d_{k+1} = x_k$, i.e., the memory will be updated. The Bellman equation may be adapted as
\begin{equation}
V^{*}_{0}(y)= \psi(x), \forall y,
\end{equation}
\begin{equation}
V^{*}_{\tau}(y)= \min_{ v\in\{0,1\} } \Big( U(x,v) + V^{*}_{\tau-1}\big(F_v(y) \big) \Big), \forall y, \forall \tau \neq 0. \label{Bellman_FinHor_ZOH_eq2}
\end{equation}
The abovementioned two equations can be used for calculating the parameters of the respective function approximators, step by step from $V^{*}_{0}(.)$ to $V^{*}_{N}(.)$, as seen earlier in Section \ref{BasicProblem}. Once the functions are approximated, the following \textit{real-time} scheduler conducts \textit{feedback scheduling} for each given $y$ and $k \in \mathbb{K}$
\begin{equation}
v_k^*(y)= \argmin_{ v\in\{0,1\} } \Big( U(x,v) + V^{*}_{N-(k+1)}\big(F_v(y) \big) \Big).\label{Scheduler_ZOH_eq1}
\end{equation}

Note that the scheduling will be conducted on the sensor side (see Fig. \ref{Fig1_NCS}). Hence, the scheduler needs to know $d_k$, that is the last successfully transmitted state information to the controller. Utilizing an \textit{acknowledgment based network communication protocol}, e.g., Transmission Control Protocol (TCP) \cite{TCP_Book}, it can be assured that the $d_k$s will be identical on both sides of the network, at each instant (of course if the acknowledgments themselves are not lost, \cite{Pappas_TAC_OptimalPOwerManagement}). The reason is the sensor will be notified whether or not the controller receives each transmitted $x_k$, in order to update its (sensor's) own copy of $d_k$ accordingly.

\section{Extension to Generalized Zero-Order-Hold} \label{GZOH_FinHor}
One applies the \textit{constant control $h(d_k)$} during the time interval in which no new sensor information is received, in ZOH. This section is aimed at using a (\textit{possibly imperfect}) model of the system on the controller side to \textit{update $d_k$} and hence the \textit{control} in the no-communication periods, called \textit{Generalized ZOH} (GZOH), \cite{Antsaklis_MB_NCS_EVentTriggering}. 
To this end, let $\hat{f}\big(.,\hat{h}(.)\big)$ be the possibly imperfect model of the plant \textit{and} its control policy, i.e., $f\big(.,h(.)\big)$. Let $\bar{k}$ denote the time at which the last state measurement was received, i.e., $d_{\bar{k}} = x_{\bar{k}}$. Updating $d_k$ may be conducted using
$d_{k+1} = \hat{f}\big(d_k,\hat{h}(d_k)\big), \forall k = \bar{k}, \bar{k}+1, ...$
when no new state information is received. Once new information is received, the variable resets to the received current value of the state. 

Considering the reason for claiming the dependency of the value function on $d_k$ in ZOH, it can be observed that if a model is used for updating $d_k$, then, the value function will depend on the \textit{current (updated) value} of $d_k$. In other words, the `overall state' of the system will include $x_k$, which is the physical state of the system, and $d_k$ which is the \textit{updated} value of $d_{\bar{k}}$, where $\bar{k}$ is the last time that the network has been scheduled. The reason for the latter dependency is the fact that the control that the controller will apply, if no new sensor information is received, will be directly dependent on the current $d_k$. 
The same relations and derivations of Section \ref{ZOH_FinHor} apply, except that Eq. (\ref{Dynamics_ZOH_y_k}) will need to be changed to
\begin{equation}
\begin{split}
y_{k+1}=F_0(y_k) := \left[ 
		\begin{array}{c}
			f\big(x_k,h(d_k)\big) \\
			\hat{f}\big(d_k,\hat{h}(d_k)\big)
		\end{array}
		\right],\\
y_{k+1}= F_1(y_k) := \left[ 
		\begin{array}{c}
			f\big(x_k,h(x_k)\big) \\
			\hat{f}\big(x_k,\hat{h}(x_k)\big)
		\end{array}
		\right].
		\label{Dynamics_GZOH_y_k}
\end{split}
\end{equation}
Note that $\hat{f}(.,.\hat{h}(.))$ will be required in both the sensor and the controller nodes, because, not only the controller needs it to update its $d_k$, but also, the sensor needs to know the current value of $d_k$ stored in the controller node. Using acknowledgment-based network communication protocols as described in Subsection \ref{ZOH_FinHor}, this condition can be fulfilled.

It is interesting to note that the idea presented in Section \ref{BasicProblem} can be simply extended to ZOH and GZOH, as done in the last two sections.
This feature shows the potential of the idea of using ADP for triggering NCSs with their different challenging issues.

\section{Extension to NCS with Random Delay and Packet Loss}

The network communications have been assume to be delay-free and loss-free, so far. The developed theories, however, can be naturally extended to the case of stochastic NCSs, e.g., networks with random delays and packet losses. The idea is utilizing the potential of the ADP/RL in handling stochastic processes \cite{Howard_PI_MDP,Sutton,Gosavi_StocasticDP}, if the \textit{probability distribution functions} of the delays and losses are known and using \textit{expected value operators} in the Bellman equation. While the details are skipped due to the page constraints, interested reader are referred to the available studies both for conventional systems \cite{Werbos2012,Bertsekas2012} and NCSs \cite{Xu_Jagannathan_Automatica}.

Besides the mentioned analytical way of handling the stochastic behaviors of the network, it is interesting to note that the \textit{feedback} scheduling nature of the presented ideas by itself is capable of handling moderate disturbances, including delays and packet losses, as shown in Section \ref{Simulations}. 

\section{Extension to Infinite-Horizon Problems}  \label{BasicProblem_InfHor}
The fact that the problems so far had a fixed and finite final time helped in developing the training algorithms. 
In other words, both having the final condition given by Eq. (\ref{Bellman_FinHor_eq1}) and the point that having $V^*_{\tau} (.)$, function $V^*_{\tau+1}(.)$ can be found using Eq. (\ref{Bellman_FinHor_eq2}) helped in developing the solution proposed in the previous sections, i.e., calculating the parameters/weights in a backward-in-time fashion. 
In many applications, however, the cost function has an \textit{infinite horizon}, e.g., regulation of states. For such a case, with a cost function similar to 
\begin{equation}
J= \sum_{k=0}^{\infty} U(x_{k},v_k), \label{CostFunction_InfHor}
\end{equation}
the objective is developing an ADP-based solution which optimizes the \textit{infinite-horizon} cost function (\ref{CostFunction_InfHor}), subject to the dynamics given in Eq. (\ref{Dynamics}) and the given control policy $h(.)$. 

Considering the infinite horizon of the problem, the concern of possible unboundedness of the cost function arises. For example, even if the network is always scheduled, a stabilizing, or even an asymptotically stabilizing, control policy $h(.)$ may lead to an unbounded cost-to-go. Therefore, motivated by the ADP literature, the definition of \textit{admissibility} is presented next and the control policy $h(.)$ is assumed to be admissible.

\begin{Def} \label{Def1}
A control policy $h(.)$ is defined to be \textit{admissible} within a compact set if a) it is a continuous function of $x$ in the set with $h(0)=0$, b) it asymptotically stabilizes the system within the set, and 
(c) the respective \textit{`cost-to-go'} or \textit{`value function'} starting from any state $x_0$ in the set, denoted with $V_h:\mathbb{R}^n \to \mathbb{R}^+$ and defined by 
\begin{equation}
V_h(x_0):=\sum_{{k}=0}^\infty {U\big(x_{k}^h, 0\big)}, \label{ValueFunction_of_h}
\end{equation}
is continuous in $\Omega$. In Eq. (\ref{ValueFunction_of_h}) one has $x_k^h:=f\big(x_{k-1}^h,h(x_{k-1}^h)\big), \forall k \in \mathbb{N}-\{0\},$ and $x_0^h := x_0$. In other words, $x_k^h$ denotes the $k$th element on the state trajectory/history initiated from $x_0$ and propagated using control policy $h(.)$. Set $\mathbb{N}$ denotes non-negative integers.
\end{Def}

The main difference between the defined admissibility and the ones typically presented in the ADP/RL literature, including \cite{AlTamimi}, is the assumption of continuity of the respective value function, instead of the assumption of its finiteness in the set. The continuity requirement is added to guarantee the possibility of \textit{uniform} approximation of the function using parametric function approximators, \cite{Weierstrass_Theorem, Hornik_NN_Continuity}. It should be noted that continuous functions are bounded in a compact set \cite{Rudin}, hence, the continuity of the value function leads to its boundedness as well, which is an essential requirement for an admissible control.

Another concern is the infinite sum over the scheduling cost, that is, the contribution of $v=1$ in the running cost of $U(x,v)$. It should be noted that the running cost of $U(x,v) = Q(x) + cv$ for a constant $c$, which was an option for fixed-final-time problems, is not desired for infinite-horizon problems. The reason is, such a cost function may become unbounded due to infinite sum of non-zero $cv_k$'s. An option for infinite-horizon cost functions is discounting such a running cost, by multiplying the running cost evaluated at $k^{th}$ time step with $\gamma^k$ for some $\gamma \in (0,1)$, which leads to the contribution of $cv$'s to be bounded by a convergent geometric series. However, it cancels the feature of utilizing the optimal value function as a Lyapunov function for proof of stability of the resulting overall system. Another option is utilizing a \textit{state-dependent scheduling weight}, which changes continuously versus the state and vanishes at the origin. For example, let $U(x,v):=Q(x)+c(x)v$, where $c:\mathbb{R}^n \to \mathbb{R}^{+}$ is a positive semi-definite function. If $c(.)$ is such that there exists a finite $\alpha\in\mathbb{R}^{+}$ which leads to $c(x) \leq \alpha Q(x), \forall x$, then, the admissibility of $h(.)$ guarantees the finiteness of (\ref{CostFunction_InfHor}).

Denoting the optimal value function of the infinite-horizon problem at state $x$ with $V^*(x)$ for some $V^*:\mathbb{R}^n \to \mathbb{R}^{+}$, the Bellman equation for such a problem reads
\begin{equation}
V^{*}(x)= \min_{ v\in\{0,1\} } \Big( U(x,v) + V^{*}\big(F_v(x) \big) \Big), \forall x, \label{Bellman_InfHor_eq1}
\end{equation}
and the optimal triggering policy is given by
\begin{equation}
v^{*}(x)= \argmin_{ v\in\{0,1\} } \Big( U(x,v) + V^{*}\big(F_v(x) \big) \Big), \forall x. \label{Bellman_InfHor_eq2}
\end{equation}

It should be noted that in infinite-horizon problems the optimal value function is not a function of time. Unlike Eq. (\ref{Bellman_FinHor_eq2}), the unknown value function exists on both sides of Eq. (\ref{Bellman_InfHor_eq1}). Therefore, instead of a recursion, one ends up with an equation to solve for the unknown value function. 
Motivated by the Value Iteration (VI) scheme in ADP/RL for solving conventional problems \cite{Sutton,Bertsekas2012}, starting with a guess on $V^0(.)$, for example $V^0(.)=0$, the iterative relation
\begin{equation}
V^{i+1}(x)= \min_{ v\in\{0,1\} } \Big( U(x,v) + V^{i}\big(F_v(x) \big) \Big), \forall x\in\Omega, \label{InfHor_Iteration}
\end{equation}
may be used for obtaining the value function of the infinite-horizon problem within compact set $\Omega$, where the superscript on $V^i(.)$ denotes the \textit{index of iteration}.
Considering the successive approximation nature of (\ref{InfHor_Iteration}), several fundamental questions arise; 1) Does the iterative relation converge? 2) What initial guess on $V^0(.)$ guarantees convergence? 3) If it converges, does it converge to the optimal value function? 4) Is the limit function, i.e., the function to which sequence $\{ V^i(.) \}_0^\infty$ converges, a continuous function, in order to use NN for \textit{uniformly} approximating it?

In a previous work of the author on the convergence analysis of ADP, these questions are answered, \cite{Heydari_TCYB, Heydari_Franklin2}. For example, it is shown that any $0\leq V^0(x) \leq U(x,0)$ leads to the convergence of the VI to the optimal value function.
The idea is establishing an \textit{analogy} between the $i$th iteration of Eq. (\ref{InfHor_Iteration}) and the time-to-go of the respective finite-horizon problem, i.e., Eq. (\ref{Bellman_FinHor_eq2}). Selecting $V^0(.) = \psi(.)$, it directly follows that $V^i(.)=V^*_i(.), \forall i$, by comparing Eq. (\ref{InfHor_Iteration}) with Eq. (\ref{Bellman_FinHor_eq2}). Hence, the convergence questions 1 to 3 simplify to whether or not the value function of the finite-horizon problem converges to the value function of the infinite-horizon problem, as the horizon extends to infinity. The answer is positive following the line of proof presented in \cite{Heydari_TCYB}. As for the fourth question, even though each $V^*_i(.)$ and hence, each $V^i(.)$ is a continuous function, it should be noted that unless the convergence of the sequence is \textit{uniform}, \cite{Rudin}, the continuity of the limit function $V^*(.)$ does not follow necessarily. 
This concern also is addressed through establishing a sufficient condition for uniform convergence of the sequence in \cite{Heydari_Franklin2}, motivated by \cite{Lincol_RelaxingDynProg, Rinehart_VI_TAC}. 

While, for simplicity, the abovementioned (infinite-horizon) results are presented for the case of NFNC, they are applicable to the more advances schemes of ZOH and GZOH as well, by utilizing $y=[x^T,d^T]^T$ as the state of the system instead of $x$, in the VI. The reason is, as the horizon extends to infinity, the fixed-final-time solutions converge to the infinite-horizon solutions. This can be seen by noting that recursive relation (\ref{Bellman_FinHor_ZOH_eq2}), applicable to both ZOH and GZOH, is actually a VI. 

\section{Stability Analysis of the Scheme for Infinite-Horizon Problems}  \label{BasicProblem_InfHor}

Some theoretical results regarding the stability of the triggering policy resulting from the VI are given in this section. The results address three different issues that will exist in almost any implementation; 1) the dynamics of the system will not be perfectly known, 2) the iterations of VI will be terminated at a finite $i$, and 3) approximation errors will exist in approximating the value functions. Before going through the analyses, an assumption needs to be made.

\begin{Assumption} \label{Assum_InvariantSet}
The intersection of the set of n-vectors $x$ at which $U(x,v) = 0$ with the invariant set of $F_v(.)$ only contains the origin, $\forall v \in \{0,1\}$. 
\end{Assumption}

Assumption \ref{Assum_InvariantSet} assures that there is no set of states (besides the set containing only the origin) in which the state trajectory can \textit{hide} forever, in the sense that the running cost evaluated at those states is zero, and hence the optimal value function is zero, without convergence of the states to the origin. 

When using ZOH and GZOH, however, it should be noted that the state vector will be $y=[x^T, d^T]^T$. If $x=0$ then for any $y=[x^T,d^T]^T$ one has $U(x,v)=0, \forall v$. Hence, especial care needs to be taken for satisfaction of Assumption \ref{Assum_InvariantSet}. For example, if $f(.,.)$ and $h(.)$ are such that $f\big(0,h(d)\big)\neq 0$ for any non-zero $d$, then, satisfaction of Assumption \ref{Assum_InvariantSet} for NFNC case leads to its satisfaction for the cases of ZOH and GZOH. The reason is, if $x_k=0$ but $d_k\neq 0$ for some $k$, then, selecting $v_k=1$, leads to $d_k=0$ and selecting $v_k=0$ leads to $y_{k} \neq F_0(y_k)$, because $x_{k+1}\neq 0$, hence, the trajectory cannot stay in $\{y=[x^T, d^T]^T\in \mathbb{R}^{2n}: x = 0, d \neq 0\}$ with $U(x,v)=0$.

\begin{Thm} \label{Stab_Model_Uncertainty}
Let the actual dynamics of the overall system, including the dynamics of the plant $f(.,.)$, the selected control policy $h(.)$, and the model of the plant and its control policy $\hat{f}\big(.,\hat{h}(.)\big)$ (if GZOH is implemented) be given by $\mathcal{F}_v:\mathbb{R}^n \to \mathbb{R}^n, v \in \{0,1\}$, where the imperfect model $F_v(.) =  \mathcal{F}_v(.) + \hat{F}_v(.)$ is used for approximating the value function through the value iteration given by (\ref{InfHor_Iteration}). If the resulting optimal value function for the imperfect model, is Lipschitz continuous with the Lipschitz constant of $\rho$, in the compact set $\Omega$, then, the resulting triggering policy $v^*(.)$ given by (\ref{Bellman_InfHor_eq2}) asymptotically stabilizes the system if
\begin{equation}
\rho \| \hat{F}_v(x) \| \leq  U(x,v), \forall x, \forall v. \label{eq1_Stab_Model_Uncertainty}
\end{equation}
\end{Thm} 

\textit{Proof}: Function $V^*(.)$, being the limit function to the recursion in (\ref{InfHor_Iteration}), satisfies (\ref{Bellman_InfHor_eq1}). By Lipschitz continuity of the value function one has 
$V^*\big(\mathcal{F}_v(x)\big) - \rho \| \hat{F}_v(x) \| \leq V^*\big(F_v(x)\big)$. Therefore,
\begin{equation}
\begin{split}
V^*\big(\mathcal{F}_{v^*(x)}(x)&\big) - V^{*}(x) \leq \\ 
&-U\big(x,v^*(x)\big) + \rho \| \hat{F}_{v^*(x)}(x) \|, \forall x. \label{eq2_Stab_Model_Uncertainty}
\end{split}
\end{equation}
If (\ref{eq1_Stab_Model_Uncertainty}) holds, Inequality (\ref{eq2_Stab_Model_Uncertainty}) leads to asymptotic stability, as $V^*(.)$ will be a Lyapunov function for $v^*(.)$. 
Note that, by Assumption \ref{Assum_InvariantSet}, no non-zero state trajectory can stay in $\{x \in \mathbb{R}^n : U(x,v) = 0, v\in\{0,1\} \}$, hence, the asymptotic stability of $v^*(.)$ follows from negative semi-definiteness of the difference between the value functions in (\ref{eq2_Stab_Model_Uncertainty}), using LaSalle's invariance theorem, \cite{Khalil}.
\qed

Assume the approximation of each value function $V^{i+1}(.)$ through (\ref{InfHor_Iteration}) leads to the approximation error of $\epsilon^i(.)$. Denoting the \textit{approximated} value function with $\hat{V}^i(.)$, it is propagated using the \textit{approximate VI} given by
\begin{equation}
\hat{V}^{i+1}(x)= \min_{ v\in\{0,1\} } \Big( U(x,v) + \hat{V}^{i}\big(F_v(x) \big) \Big) + \epsilon^i(x), \forall x. \label{eq1_AVI}
\end{equation}
Once the approximation errors exist, the convergence of the value functions to the optimal value function is no longer guaranteed. As a matter of fact, it is not even guaranteed that the iterations converge. Following the line of proof in \cite{Heydari_SAVI}, it can be proved that if $|\epsilon^i(x)| \leq c U(x,0), \forall x, \forall i$, for some $c\in [0,1),$ then each $\hat{V}^i(.)$ is lower and upper bounded, respectively, by $\underline{V}^i(.)$ and $\overline{V}^i(.)$ at each $i$, if $\underline{V}^0(x) \leq \hat{V}^0(x) \leq \overline{V}^0(x), \forall x,$ where, 
\begin{equation}
\underline{V}^{i+1}(x)= \min_{ v\in\{0,1\} } \Big( U(x,v) - cU(x,0) + \underline{V}^i\big(F_v(x) \big) \Big), \forall x, \label{eq1_AVI_Lowerbound_VI}
\end{equation}
\begin{equation}
\overline{V}^{i+1}(x)= \min_{ v\in\{0,1\} } \Big( U(x,v) + cU(x,0) + \overline{V}^i\big(F_v(x) \big) \Big), \forall x. \label{eq1_AVI_Upperbound_VI}
\end{equation}
In other words, $\underline{V}^i(.)$ and $\overline{V}^i(.)$ correspond to the \textit{exact} value iteration for cost functions
\begin{equation}
\underline{J}= \sum_{k=0}^{\infty} \big(U(x_{k},v_k) - cU(x_{k},0)\big), \label{eq1_AVI_Lowerbound_CostFUnction}
\end{equation}
\begin{equation}
\overline{J}= \sum_{k=0}^{\infty} \big(U(x_{k},v_k) + cU(x_{k},0)\big). \label{eq1_AVI_Lowerbound_CostFUnction}
\end{equation}
The next theorem provides a sufficient condition for stability of both having an approximation error and/or terminating the iterations after a finite $i$.
\begin{Thm} \label{Stab_Approx_Errors}
Let the approximate value iteration given by (\ref{eq1_AVI}) be conducted using a continuous function approximator with the approximation error bounded by $|\epsilon^i(x)| \leq c U(x,0), \forall x, \forall i$, for some $c\in [0,1).$ Let the iteration terminate at the $i^{th}$ iteration when 
\begin{equation}
|\hat{V}^i(x) - \hat{V}^{i+1}(x) | \leq \delta(x), \forall x\in\Omega, \label{eq1_AVI_Lowerbound_CostFUnction}
\end{equation}
for some positive (semi-)definite real-valued function $\delta(.)$. The resulting triggering policy $\hat{v}^i(x) = \argmin_{ v\in\{0,1\} } \big( U(x,v) + \hat{V}^{i}\big(F_v(x) \big) \big)$ asymptotically stabilizes the system if
\begin{equation}
\delta(x) <  (1-c) U(x,0), \forall x\in\Omega-\{0\}. \label{eq1_Stab_Approx_Errors}
\end{equation}
\end{Thm} 

\textit{Proof}: By (\ref{eq1_AVI}) and (\ref{eq1_AVI_Lowerbound_CostFUnction}) one has
\begin{equation}
\hat{V}^{i}(x) + \delta(x) \geq U\big(x,\hat{v}^i(x)\big) + \hat{V}^i\big(F_{\hat{v}^i(x)}(x)\big) + \epsilon^i(x), \forall x. \label{eq2_Stab_Approx_Errors}
\end{equation}
or
\begin{equation}
\hat{V}^i\big(F_{\hat{v}^i(x)}(x)\big) - \hat{V}^{i}(x) \leq - U\big(x,\hat{v}^i(x)\big) - \epsilon^i(x) + \delta(x), \forall x. \label{eq3_Stab_Approx_Errors}
\end{equation}
Considering $|\epsilon^i(x)| \leq c U(x,0)$ and (\ref{eq1_Stab_Approx_Errors}), the foregoing inequality along with Assumption \ref{Assum_InvariantSet} lead to the asymptotic stability of the system. Note that the lower boundedness of $\hat{V}^i(.)$ by $\underline{V}^i(.)$ guarantees the positive definiteness of the candidate Lyapunov function, and its continuity follows from the continuity of the selected function approximator.
\qed

Considering the point that the approximation errors exist, inequality (\ref{eq1_AVI_Lowerbound_CostFUnction}) may never be satisfied. In this case, one needs to decrease the approximation error, through selecting a richer function approximator. Such an action is helpful, because of the lower and upper boundedness of $\hat{V}^i(.)$, by the exact value functions and the convergence of the exact value functions to the same optimal value function as $c \to 0$ and $i \to \infty$.

\section{Extension to Model Free Schemes} \label{Model_Free}
All the previously discussed ideas require the knowledge of the dynamics of the plant. For example, considering the simple case of NFNC, Section \ref{BasicProblem}, the model will be required in two places. 1) In offline training, because of the existence of $F_v(.)$ in Eq. (\ref{Bellman_FinHor_eq2}). 2) In online scheduling, because of the existence of $F_v(.)$ in Eq. (\ref{Scheduler_eq1}). The objective in this section is developing completely \textit{model free} methods. An idea toward this goal is utilizing two separate concepts from the ADP/RL literature in control of conventional problems; a) learning action-dependent value functions \cite{Werbos_ActionDependent_Critic} which is called Q-function in Q-learning \cite{Watkins} and b) conducting online learning \cite{Werbos2012,AlTamimi}. 

Consider the infinite-horizon problem discussed in Section \ref{BasicProblem_InfHor}. 
Let the \textit{action-dependent value function} $V^*(x,v)$ denote the incurred cost if \textit{action} $v$ is taken at the current time and the \textit{optimal actions} are taken for the future times, for some $V:\mathbb{R}^n \times \{0,1\} \to \mathbb{R}^+$. Then, by definition, one has
\begin{equation}
V^{*}(x,v)=  U(x,v) + \min_{ w\in\{0,1\} } V^{*}\big(F_v(x),w \big), \forall x, \forall v. \label{Bellman_ActionDep_eq2}
\end{equation}
Motivated by VI and the previous section, the solution to (\ref{Bellman_ActionDep_eq2}) can be obtained through selecting an initial guess $V^0(.,.)$ and conducting the successive approximation given by
\begin{equation}
V^{i+1}(x,v)=  U(x,v) + \min_{ w\in\{0,1\} } V^{i}\big(F_v(x),w \big), \forall x, \forall v. \label{eq_ADVI}
\end{equation}
This learning process is called Q-Learning, \cite{Watkins, Werbos_ActionDependent_Critic}.

For simplicity, the idea of NFNC is being considered here for deriving a model free algorithm; however, other schemes, e.g., ZOH, directly follow, by replacing $x$ with $y$. 
The interesting feature of action dependent value functions is the fact that the scheduler does not require the model of the system, since,
\begin{equation}
v^*_k(x_k)= \argmin_{ v\in\{0,1\} } V^{*}(x_k,v), \forall x_k. \label{Scheduler_ActionDep_eq1}
\end{equation}
Note that, $V^*(x) = V^*\big(x, v^*(x) \big)$, where $V^*(x)$ is the (action independent) value function presented in Section \ref{BasicProblem_InfHor} which satisfies Eq. (\ref{Bellman_InfHor_eq1}).
Therefore, using this idea, the need for the model in the scheduling stage can be eliminated. As for the need for the model in the training, i.e., in Eq. (\ref{eq_ADVI}), the following idea can be utilized. If the learning is conducted `on the fly', i.e., \textit{online learning}, then, no model of the system is required for training also. The reason is, instead of using $f(x,u)$ for finding $x_{k+1}$ to be used in the right hand side of Eq. (\ref{eq_ADVI}), one can wait for one time step and measure $x_{k+1}$ directly.
As a matter of fact, this is how we learn, for example, to drive a car, i.e., by waiting and observing the outcomes of the taken actions, \cite{Watkins,Sutton,Werbos2012}. 

Let $V^*(.,0)$ and $V^*(.,1)$ be approximated using two separate function approximators (because of the possible discontinuity of $V(.,.)$ with respect to its second argument). Considering Eq. (\ref{eq_ADVI}), the outline of the online learning process is presented in Algorithm 1.

\vspace{10pt}
\textbf{Algorithm 1}

Step 1: Select initial guesses $V^0(.,0)$ and $V^0(.,1)$ and set $k=0$. 

Step 2: Randomly select $v_k\in\{0,1\}$.

Step 3: Apply $v_k$ on the system, i.e., trigger or don't trigger based on the selected $v_k$.

Step 4: Wait for one time step and measure $x_{k+1}$.

Step 5: Update the action dependent value function corresponding to the selected $v_k$ using
\begin{equation}
V^{k+1}(x_k,v_k)=  U(x_k,v_k) + \min_{ w\in\{0,1\} } V^{k}\big(x_{k+1},w \big), \label{InfHor_ActionDep_Iteration}
\end{equation}
and keep the value function of the action which was not taken constant, i.e., $$V^{k+1}(x_k,v) = V^{k}(x_k,v), \mbox{ for } v\neq v_k.$$

Step 6: Go back to Step 2.

\vspace{10pt}

An important point is, the proposed scheme is entirely model-free and does not need any \textit{model identification}, unlike the (model-free) methods that conduct a separate model identification phase to identify the model and then use the result in the learning or control process, e.g. \cite{Sahoo_Jagannathan_ACC13}.

It is worth mentioning that the presented algorithm fits in the category of \textit{exploration} in the RL literature, \cite{Sutton}. This is equivalent of the condition of persistency of excitation \cite{AlTamimi} in conventional optimal and adaptive control. The point is, through the random selection of the decisions, the algorithm gives the chance to the parameters of every function approximator to learn different `behaviors' of the system.

\section{Stability of Model-Free Schemes} \label{Stable_Mode_Free}
Section \ref{Model_Free} presented the idea for a model-free scheduler, which calls for {\textit{online learning}}. However, the \textit{stability} of the system during the online learning can be at stake, considering the point that the decisions are made randomly. Motivated by ADP/RL, one may conduct \textit{exploitation} as well, through replacing Step 2 of Algorithm 1 with the following step, in some iterations.

Step 2: Select $v_k = \argmin_{ v\in\{0,1\} } V^{k}(x_k,v)$. \\
It should be noted that one will still need to do exploration occasionally, through selecting different random $v_k$'s. A reason is, if a particular $v$ is not selected in Step 2 of the exploitation algorithm, the respective value function will never get the chance to be learned, i.e., the parameters of the function approximator corresponding to that action will never be updated in Step 5.
While the selection of the decisions through the exploitation phase decreases the concern (as compared with the random selection of the decisions), the stability concern still exists. The reason is, the decisions made in the exploitation actions are based on the current, possibly immature, version of the value function,
\begin{equation}
v^{i}(x) =  \argmin_{ w\in\{0,1\} } V^{i}(x,w), \forall x\in \Omega, \forall i. \label{eq_ADVI_vi}
\end{equation}
As long as $V^{i}(.,.)$ is not optimal, its resulting policy, $v^i(.)$, may not even be stabilizing. This concern can be addressed through utilizing the value function of a stabilizing triggering policy as the initial guess, presented in a previous work in \cite{Heydari_SAVI}. Ref. \cite{Heydari_SAVI}, however, investigated conventional optimal control problems (where the decision variable changes continuously) with an action-independent value function. The rest of this section is devoted to adapting the stability and convergence results presented in that work to both the decision making/switching case and the case of action dependence, i.e., the problem at hand. The adaptation of the results is not trivial, hence, the proofs of the theorems, along with some required lemmas, are included in the Appendex of this paper. These results can also be extended to the case of having approximation errors, using the derivations in \cite{Heydari_SAVI} as the basis (skipped due to the page constraints).

Considering the definition of the (action independent) value function of a \textit{control} policy $h(.)$ assuming constant access to state measurement, given in Definition \ref{Def1} and denoted with $V_h(.)$, the action dependent value function of a \textit{triggering} policy given by $w(.)$ may be defined as 
\begin{equation}
\begin{split}
V_{w}(x_0,v) := U(x_0,v) + \sum_{k=1}^{\infty} &{U\big(x^w_{k}, w(x^w_{k})\big)}, \\ 
&\forall x_0 \in \Omega, \forall v \in \{0,1\}, \label{eq_V0_Definition}
\end{split}
\end{equation}
where $x_k^w:=F_{w(x_{k-1})}(x_{k-1}), \forall k \in \mathbb{N}-\{1, 0\},$ and $x_1^w := F_v(x_0)$. In other words, $x_k^w$ denotes the $k$th element on the state trajectory initiated from $x_0$ and propagated using triggering decision $v$ for the first time step and then using triggering policy $w(.)$ for the rest of the times. Obviously $V_{w}(.,.)$ depends on the control policy $h(.)$ as well, but, as long as it is clear, the inclusion of `$h$' is skipped in the notation for the value function, for notational simplicity. 
Considering (\ref{eq_V0_Definition}), it can be seen that $V_{w}(.,.)$ solves
\begin{equation}
		V_w(x,v) = U(x,v) + V_w\Big(F_v(x),w\big(F_v(x)\big)\Big), \forall x, \forall v. \label{eq_ADVI_V0}
\end{equation}

\begin{Def} \label{Admissible_Triggering_Policy}
The triggering policy $w(.)$ whose action dependent value function $V_w(.,v)$ is continuous and hence bounded in a compact set, $\forall v \in \{0,1\}$, is called an admissible triggering policy.
\end{Def} 

\begin{Def} \label{StabilizingADVI_Definition}
The action-dependent value iteration (ADVI) scheme given by recursive relation (\ref{eq_ADVI}) which is initiated using the action dependent value function of an admissible triggering policy is called stabilizing action dependent value iteration (SADVI).
\end{Def}

Besides the theoretical stability analyses, presented next, for practice however, how can one find an initial admissible triggering policy, especially if the dynamics of the system is not known? To address this question, the fact that we already have an admissible control policy $h(.)$ on the controller side should be considered. The simple triggering policy of $w(x)=1, \forall x$ is an admissible triggering policy. Once an admissible triggering policy is selected, its action dependent value function can be obtained using Theorem \ref{Thm_Conv_V0}.

Let $V(.,v) \in \mathcal{C}(x)$ (respectively, $V(.,.) \in \mathcal{C}(\Omega\times\{0,1\})$) denote that function $V(.,v)$ is continuous at point ${x}$ for the given $v\in\{0,1\}$, (respectively, within $\Omega$ for every given $v$).

\begin{Thm} \label{Thm_Conv_V0} If $w(.)$ is an admissible triggering policy within $\Omega$, then selecting any $V_{w}^{0}(.,.)\in \mathcal{C}(\Omega\times\{0,1\})$ which satisfies $0 \leq V_{w}^{0}(x,v) \leq U(x,v), \forall x \in \Omega, \forall v \in \{0,1\}$, the iterations given by
\begin{equation}
		V_{w}^{i+1}(x,v) = U(x,v) + V_{w}^{i}\Big(F_v(x),w\big(F_v(x)\big) \Big), \forall x \in \Omega, \label{PI_Value_eq2}
\end{equation}
converge uniformly to the action dependent value function of $w(.)$, in compact set $\Omega$. 
\end{Thm}

The next theorem proves the convergence of SADVI to the optimal action dependent value function.

\begin{Thm} \label{Theorem_VI_Convergence} The stabilizing action dependent value iteration converges to the optimal action dependent value function of the infinite-horizon problem within the selected compact domain.
\end{Thm}

\begin{Thm} \label{Thm_Stab_LyapFun}
Let the compact domain $\mathcal{B}^{i}_r$ for any $r\in\mathbb{R}^+$ be defined as $\mathcal{B}^{i}_r:=\{ x\in\mathbb{R}^n : {V}^{i}\big(x,v^i(x)\big) \leq r \}$ and let $\bar{r}^i > 0$ be (the largest $r$) such that $\mathcal{B}^{i}_{\bar{r}^i} \subset \Omega$. Then, for every given $i \in \mathbb{N}$, triggering policy $v^{i}(.)$ resulting from stabilizing action dependent value iteration asymptotically stabilizes the system about the origin and $\mathcal{B}^{i}_{\bar{r}^i}$ will be an estimation of the region of attraction for the system.
\end{Thm}

Theorem \ref{Thm_Stab_LyapFun} proves that each single $v^i(.)$ if \textit{constantly} applied on the system, will have the states converge to the origin. However, in online learning, the triggering policy will be subject to adaptation. In other words, if $v^i(.)$ is applied at the current time, triggering policy $v^{i+1}(.)$ will be applied at the next time-step. It is important to note that even though Theorem \ref{Thm_Stab_LyapFun} proves the asymptotic stability of the \textit{autonomous} system $x_{k+1}=F(x_k):=F_{v^{i}(x_k)}(x_k)$ for every fixed $i$, it does not guarantee the asymptotic stability of the \textit{time varying} system $x_{k+1}=F(x_k,k):=F_{v^{k}(x_k)}(x_k)$.
Therefore, following a similar discussion in \cite{Heydari_SAVI} for conventional problems, it is required to have a separate stability analysis to show that the trajectory formed under the \textit{adapting/evolving} triggering policy also will converge to zero. An idea for doing that is finding a \textit{single} function to be a Lyapunov function for \textit{all} the triggering policies. The proof of the following theorem, however, uses another approach. 

\begin{Thm} \label{Thm_Stabil_SADVI_No_Lyap} 
If the system is operated using triggering policy $v^{k}(.)$ at time $k$, that is, the policy subject to adaptation in the stabilizing action dependent value iteration, then, the origin will be asymptotically stable and every trajectory contained in $\Omega$ will converge to the origin.
\end{Thm}

Finally, the next theorem provides an idea for finding an estimation of the region of attraction (EROA) for the evolving triggering policy.

\begin{Thm} \label{Thm_ROA_Evolving_VI} 
Let $\mathcal{B}^{i}_r:=\{ x\in\mathbb{R}^n : {V}^{i}\big(x,v^i(x)\big) \leq r \}$ and $\mathcal{B}^*_r:=\{ x\in\mathbb{R}^n : {V}^*\big(x,v^*(x)\big) \leq r \}$ for any $r\in\mathbb{R}^+$. Also, let the system be operated using triggering policy $v^{k}(.)$ at time $k$, that is, the control subject to adaptation in the stabilizing value iterations. 
If $\mathcal{B}^{*}_{r} \subset \Omega$ for an $r>0$ then $\mathcal{B}^{0}_{r}$ is an estimation of the region of attraction of the closed loop system.
\end{Thm}

Before concluding the theoretical analyses on the stability of the system operated under SADVI (online learning), it should be added that Algorithm 1 updates $V^{k+1}(.,.)$ only at the current state at each time. But, the presented theory on the stability of the system under online learning is based on (\ref{eq_ADVI}), i.e., each new value function is calculated based on the entire $\Omega$. A possible scenario for satisfying such a condition is having multiple similar sensor-controller-plant sets evolving together and sharing their `observations' at different respective states with each other, to end up with one new $V^{k+1}(.,.)$ at each time step. If this condition is not satisfied, one may still utilize the \textit{pointwise} update given by Algorithm 1, however, the stability will then depend on the generalization capability of the function approximator and is not directly guaranteed.
In this case, an idea is monitoring the states to switch to an available stabilizing triggering policy within $\Omega$, (e.g., $w(x)=1, \forall x$) if the states approached the boundaries of $\Omega$, which is an EROA for the stabilizing policy. However, once the learning converges for different states within $\Omega$, the stability of the proposed scheme is guaranteed by Theorem \ref{Thm_Stab_LyapFun}, with the respective EROA.

\section{Numerical Examples} \label{Simulations}
In order to demonstrate the potential of the schemes in practice, a few simple examples are simulated in this section.
\subsection{ZOH}
\begin{figure}[t]
  \vspace{-10pt}
  \begin{center}
		\includegraphics[width=0.95\columnwidth]{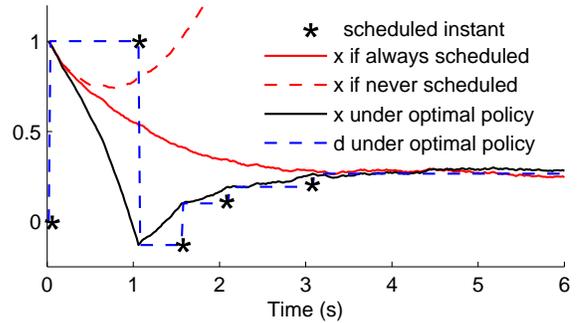}
  \end{center}
  \vspace{-15pt}
		\caption{Simulation results for the ZOH scheme under a random disturbance term acting on the plant.} 
		\label{FigSim1}
  \vspace{-10pt}
\end{figure}

The simulated plant is a scalar system with the dynamics of $\dot{x} = \sin(x)+u$ and the selected control policy is $h(x) = -\sin(x)-x$. The problem is discretized with sampling time $20$ $ms$ and $N=300$. The ZOH scheme is implemented using $Q(x) = x^2$, $c = 0.01$, $\psi(x)=25x^2$. 
The function approximator was selected in a polynomial form made of $x$ and $d$, up to the fourth order, where the coefficients are the tunable parameters.
The approximation domain was selected as $\Omega:=\{y=[x,d]^T\in [-2,2] \times [-2,2]\}\subset \mathbb{R}^2$. 100 new random $y$s were selected from $\Omega$ in each evaluation of (\ref{Bellman_FinHor_ZOH_eq2}) to conduct least squares for finding the parameters. The approximation of the value functions took less than 3 seconds in a computer with the CPU of Intel Core i7, 3.4 GHz running MATLAB in single threading mode. The result was utilized for controlling initial condition $x_0 = 1$. 
In order to simulate real-world conditions, a random time-varying disturbance force uniformly distributed between $0$ and $0.01$ was applied on the system through its control input. The disturbance is selected large enough to destabilize the system when operated in an open loop fashion, that is, when the control is calculated ahead of time, from an assumed disturbance-free state trajectory resulting from $\dot{x} = \sin(x) + h(x) = -x$. The simulation results, as presented in Fig. \ref{FigSim1}, show that the scheduler has been able to control the state of the system through only 5 network transmissions. For comparison purposes, the state history if the network is always scheduled and the state history if it is operated in the described open loop fashion are also plotted. 
The proposed method calls for the communication load which is less than 2\% of what the always scheduled case requires.

\subsection{GZOH with Random Packet Losses}
\begin{figure}[t]
  \vspace{-10pt}
  \begin{center}
		\includegraphics[width=0.95\columnwidth]{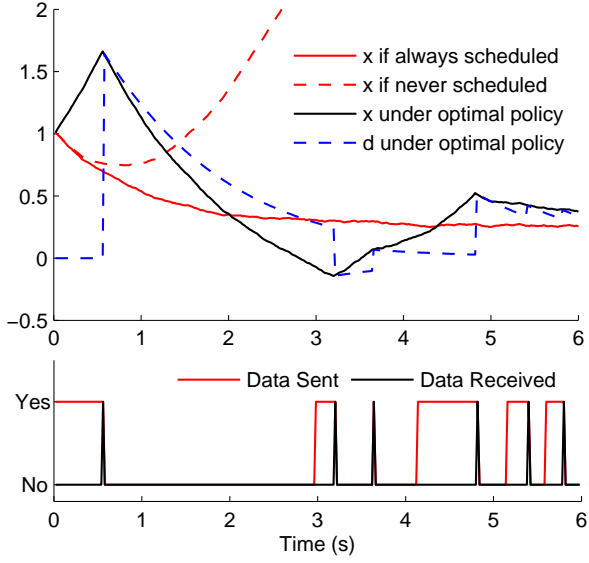}
  \end{center}
  \vspace{-10pt}
		\caption{Simulation results for the GZOH scheme under a random disturbance term acting on the plant and the packet loss probability of 0.9.} 
		\label{FigSim2}
  \vspace{-5pt}
\end{figure}

To see the performance of the method in GZOH and also in dealing with lossy networks even without incorporating the stochastic nature of the problem with random losses as in \cite{Xu_Jagannathan_Automatica}, the previous example is modified as follows. The imperfect model $\dot{x}= -0.25 \sin(x) + u$ and imperfect control policy $\hat{h}(x) = -0.5x$, instead of the actual system and control policy, are utilized for GZOH. Moreover, it is assumed that the transmitted packets will be dropped with a 90\% chance. The random disturbance force of the previous simulation is also applied. The results, presented in Fig. \ref{FigSim2}, show the capability 
of the controller in controlling the state and dealing with the very high packet loss probability. As seen in the history of data transmission, when the scheduler tries to send a state measurement to the controller if it fails, the scheduler \textit{keeps trying} to send until it is successful. Another important feature of the result is utilization of the GZOH, which lead to updating $d_k$ in Fig. \ref{FigSim2}, instead of keeping it constant as in Fig. \ref{FigSim1}.

\subsection{Model Free GZOH}
In order to simulate the performance of the model-free scheme, the dynamics of Van der Pol's oscillator, $\ddot{z} = (1-z^2)\dot{z} - z +u$, and the (feedback linearization based) policy $h(z) = -(1-z^2)\dot{z} + z - 2z -2\dot{z}$ was chosen. The problem was taken into state space by defining $x=[x_1, x_2]^T:=[z,\dot{z}]^T$ and discretized with sampling time $25$ $ms$. The GZOH scheme was implemented using $Q(x) = 0.625x^Tx$ and $c(x) = 2Q(x)$ with the imperfect model $\ddot{z} = u$ and imprecise control policy $\hat{h}(z) = -z -\dot{z}$. 
The function approximator was selected in a polynomial form made of elements of $x$ and $d=[d_1, d_2]^T$, up to the third order. 
The approximation domain was selected as $\Omega:=\{y=[x,d]^T: x,d\in [-2,2] \times [-2,2]\} \subset \mathbb{R}^4$. The action dependent value function of $h(.)$ was calculated, using Theorem \ref{Thm_Conv_V0}, as the initial guess for our online SADVI.

\begin{figure}[t]
  \vspace{-10pt}
  \begin{center}
		\includegraphics[width=0.95\columnwidth]{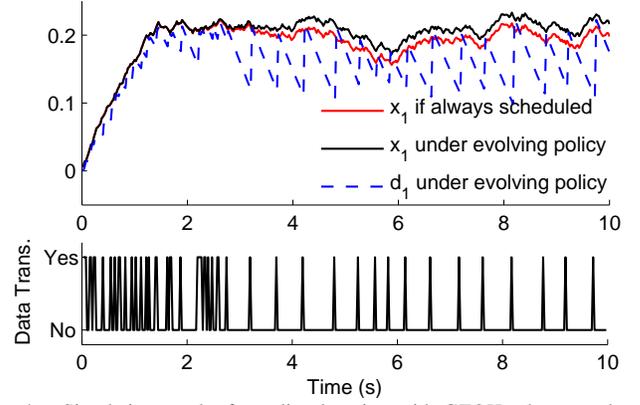}
  \end{center}
  \vspace{-10pt}
		\caption{Simulation results for online learning with GZOH scheme under a random disturbance term.} 
		\label{FigSim3}
  \vspace{-5pt}
\end{figure}

A disturbance, uniformly distributed between $0$ and $0.01$ and acting on $x_1$ as an additive term, was applied, which destabilizes the system in case of no feedback information. Using $w(x)=1, \forall x,$ the system initially at the origin, will be stabilized, as shown by the red plots in Fig. \ref{FigSim3}. This policy however requires network communications for the entire time. Using online learning through SADVI, \textit{without} using the dynamics of the system, the control policy, or their respective models used in GZOH, the value function was updated within the first 3 seconds through both the exploration and exploitation algorithms (chosen randomly at each time step). Afterward, the updated triggering policy was utilized for scheduling the network. The resulting state trajectory (including the first 3 seconds of learning) is depicted in Fig. \ref{FigSim3}. by the black plot. The history of the respective $d_1$ is also plotted. As seen, after the end of the learning, the network communication is decreased to a fraction of the times. This leads to a considerable saving of the network bandwidth, compared with the initial admissible triggering policy. 

\section{Conclusions}
ADP was shown to be very promising in designing triggering policies which provide (near) optimal solutions to the networked control problems. The approach is very versatile in extending to ZOH, GZOH, stochastic networks, infinite-horizon problems, and problems with unknown/uncertain dynamics. These features along with the low realtime computational load of the scheme make it very desirable. However, the approach calls for more theoretical rigor, for proof of stability and optimality. While this study took several steps to this end, many questions are left to be answered, particularly for stability of systems during online learning and the generalization capability of the function approximators.

\appendix
This Appendix includes the proofs of Theorems \ref{Thm_Conv_V0} to \ref{Thm_ROA_Evolving_VI}.

\textit{\textbf{Proof of Theorem \ref{Thm_Conv_V0}}}: 
Eq. (\ref{PI_Value_eq2}) leads to
\begin{equation}
V_{w}^{i}(x_0,v)= U(x_0,v) + \sum_{k=1}^{i-1} {U\big(x^w_{k}, w(x^w_{k})\big)} + V_{h}^{0}\big(x^w_{i},w(x^w_{i})\big). \label{V0_Conv_Lem_eq1}
\end{equation}
Comparing (\ref{V0_Conv_Lem_eq1}) with (\ref{eq_V0_Definition}) and considering $0 \leq V_{w}^{0}(x,v) \leq U(x,v)$ one has $0 \leq V_{w}^{i}(x,v) \leq V_w(x,v)$. Therefore, sequence $\{ V_w^i(x,v) \}_{i=0}^{\infty}$ is upper bounded by $V_w(x,v)$, for each given $x$ and $v$. The limit function $V_w^{\infty}(.,.)$ is equal to $V_w(.,.)$, since, the admissibility of $w(.)$ leads to $x_{i}^w \to 0$, and hence, $V_{w}^{0}(x_{i}^h,v) \to 0$ as $i \to \infty$, due to $0 \leq V_{w}^{0}(x,v) \leq U(x,v)$. Hence, (\ref{V0_Conv_Lem_eq1}) converges to (\ref{eq_V0_Definition}) as $i \to \infty$. This proves \textit{pointwise} convergence of the sequence to $V_w(.,.)$.

But, this convergence is monotonic. Note that for any arbitrary positive integers $i_1$ and $i_2$, if $i_1 \leq i_2$, then
\begin{equation}
\begin{split}
V&_{w}^{i_1}(x_0,v) - V_{w}^{i_2}(x_0,v)= V_{w}^{0}\big(x^w_{i_1},w(x^w_{i_1})\big) -\\
&V_{w}^{0}\big(x^w_{i_2},w(x^w_{i_2})\big) - \sum_{k=i_1}^{i_2-1} {U\big(x^w_{k}, w(x^w_{k})\big)} \leq 0, \forall x, \forall v,  \label{PI_Value_eq3_2}
\end{split}
\end{equation}
since $0\leq V_{w}^{0}\big(x^w_{i_1},w(x^w_{i_1})\big)\leq U\big(x^w_{i_1},w(x^w_{i_1})\big)$, and the last term in the foregoing inequality is only one of the non-negative terms in the summation in the right hand side of (\ref{PI_Value_eq3_2}). Therefore, sequence of functions $\{V_{w}^{i}(x,v)\}_{i=0}^{\infty}$ is pointwise non-decreasing. On the other hand, the limit function is continuous, by admissibility of the triggering policy, and also, each element of the sequence is continuous, as it is a finite sum of continuous functions. These characteristics lead to uniform convergence of the sequence in the compact set, by Dini's uniform convergence theorem (Ref. \cite{Rudin}, Theorem 7.13).
\qed

The following lemma establishes a monotonicity feature to be used in the proof of Theorem \ref{Theorem_VI_Convergence}.
\begin{Lem} \label{Lemma_NonDecreasing} 
Sequence of functions $\{ V^{i}(x,v) \}_{i=0}^{\infty} := \{V^{0}(x,v), V^{1}(x,v), ... \}$ resulting from stabilizing action dependent value iteration is a pointwise non-increasing sequence.
\end{Lem}
\textit{\textbf{Proof of Lemma \ref{Lemma_NonDecreasing}}}: 
The proof is done by induction. Considering (\ref{eq_ADVI_V0}), which gives $V^0(.,.)$, and (\ref{eq_ADVI}), which (for $i=0$) gives $V^1(.,.)$, one has 
\begin{equation}
		V^1(x,v) \leq V^0(x,v), \forall x \in \Omega, \forall v \in \{0,1\}, \label{Lemma1_eq1}
\end{equation}
because $V^1(.,.)$ is the result of minimization of the right hand side of (\ref{eq_ADVI}) instead of being resulted from a given triggering policy $w(.)$.
Now, assume that for some $i$, we have
\begin{equation}
		V^i(x,v) \leq V^{i-1}(x,v), \forall x \in \Omega, \forall v \in \{0,1\}. \label{Lemma1_eq2}
\end{equation}
Considering the definition of $v^{i-1}(.)$ given by (\ref{eq_ADVI_vi}) and using (\ref{Lemma1_eq2}) as well as the fact that $v^{i-1}(x)$ is not necessarily the minimizer used in the definition of $V^{i+1}(.,.)$, lead to 
\begin{equation}
		\begin{split}
		V^{i+1}(x,v) = U(x,v) + \min_{ w\in\{0,1\} } V^{i}\big(F_v(x), w \big) \\
		\leq U(x,v) + V^{i}\Big(F_v(x),v^{i-1}\big(F_v(x)\big) \Big) \\
		\leq U(x,v) + V^{i-1}\Big(F_v(x),v^{i-1}\big(F_v(x)\big) \Big) = V^{i}(x,v), \\
		\forall x \in \Omega, \forall v \in \{0,1\}. \label{Lemma1_eq4}
		\end{split}
\end{equation}
Therefore, $V^{i+1}(x,v) \leq V^{i}(x,v)$ which completes the proof.
\qed

\textit{\textbf{Proof of Theorem \ref{Theorem_VI_Convergence}}}: 
Considering finite-horizon cost function (\ref{CostFunction_FinHor}), the action dependent optimal value function $V^{*}_{N}(x,v)$ is defined as the cost of taking action $v$ at the first time step and taking the (history of) time-dependent optimal actions with respect to the cost function with the horizon of $N-1$ for the remaining time steps. In other words, $\min_{v} V^{*}_{N}(x,v) = V^{*}_{N}(x)$. Therefore,
\begin{equation}
V^{*}_1(x,v)= U(x,v) + \psi\big(F_v(x)\big), \forall x, \label{AD_Bellman_FinHor_eq1}
\end{equation}
\begin{equation}
\begin{split}
V^{*}_{\tau}(x,v)= U(x,v) + \min_{ w\in\{0,1\} } V^{*}_{\tau-1}\big(F_v(x),w \big), \\
\forall x, \forall v, \forall \tau =2,3,.... \label{AD_Bellman_FinHor_eq2}
\end{split}
\end{equation}

Selecting $\psi(x)$ such that $V^0 (x,v) = U(x,v) + \psi\big(F_v(x)\big)$, one has
\begin{equation}
V^{i-1}(x,v) = V^*_i(x,v), \forall x, \forall v, \forall i \in \mathbb{N}-\{0\}. \label{AD_Bellman_FinHor_eq3}
\end{equation}
The foregoing equation corresponds to the analogy between the finite horizon and infinite horizon action-independent value functions, discussed earlier and detailed in \cite{Heydari_TCYB}.

By the non-increasing (cf. Lemma \ref{Lemma_NonDecreasing}) and non-negative (by definition) nature of value functions under SADVI, they converge to some limit function $V^{\infty}(.,.)$. Considering (\ref{AD_Bellman_FinHor_eq3}), the limit function is the optimal action dependent value function to cost function (\ref{CostFunction_InfHor}). 
This can be observed by noticing that due to the convergence of SADVI, one has 
$\lim_{i \to \infty} x_i \to 0$
using decision sequence $\{v, v^*_{1}(x_1), v^*_{2}(x_2), ..., v^*_{i-1}(x_{i-1})\}$ which is the sequence of decisions taken in evaluating cost-to-go $V^*_i(x,v)$. Otherwise, $V^{\infty}(.,v)$ becomes unbounded. 
Note that per Assumption \ref{Assum_InvariantSet} the state trajectory cannot hide in the invariant set of $F_v(.)$ with zero cost, to lead to a finite cost-to-go without convergence to the origin.
Therefore, by $\lim_{i \to \infty} x_i \to 0$ one has
\begin{equation}
\lim_{i \to \infty} \psi(x_i) \to 0, \label{Thm2_eq2}
\end{equation}
in calculation of ${V}^*_i(.,.)$. Comparing finite-horizon cost function (\ref{CostFunction_FinHor}) with infinite-horizon cost function (\ref{CostFunction_InfHor}) and considering (\ref{Thm2_eq2}), one has
\begin{equation}
V^*(x,v)=V^{\infty}(x,v), \forall x \in \Omega, \forall v \in \{0,1\}. \label{Thm2_eq3}
\end{equation}
Otherwise, the smaller value among $V^*(x,v)$ and $V^{\infty}(x,v)$ will be both the optimal action dependent value function (evaluated at $x$) for the infinite-horizon problem and the greatest lower bound of the sequence of value function of the fixed-final-time problems resulting from $N = 0, 1, 2, ...$.
\qed

The next lemma proves the continuity of each action dependent value function resulting from SADVI, to be used for stability analysis, in proof of Theorem \ref{Thm_Stab_LyapFun}.
\begin{Lem} \label{Lemma_Continuity_SADVI} 
Each element of the sequence of functions $\{ V^{i}(x,v) \}_{i=0}^{\infty} := \{V^{0}(x,v), V^{1}(x,v), ... \}$ resulting from stabilizing action dependent value iteration is a continuous function of $x, \forall v \in \{0,1\}$.
\end{Lem}
\textit{\textbf{Proof of Lemma \ref{Lemma_Continuity_SADVI}}}: 
The continuity of each action-independent value function $V^i(.)$ initiated from a continuous initial guess and generated using (\ref{InfHor_Iteration}) for the general case of switching problems is proved in \cite{Heydari_Franklin2}. 
Considering (\ref{eq_ADVI_vi}) and comparing (\ref{eq_ADVI}) with (\ref{InfHor_Iteration}) one has 
\begin{equation}
V^i\big(.,v^i(.)\big) = V^i(.), \forall i. \label{eq_ADVI_vs_VI}
\end{equation}
From $V^i(.) \in \mathcal{C}(\Omega)$ for every finite $i$, $U(.,.) \in \mathcal{C}(\Omega\times\{0,1\})$, $F_v(.) \in \mathcal{C}(\Omega), \forall v$, and Eq. (\ref{eq_ADVI_vs_VI}), it follows that $V^i(.,.) \in \mathcal{C}(\Omega\times\{0,1\})$ by Eq. (\ref{eq_ADVI}).
\qed

\textit{\textbf{Proof of Theorem \ref{Thm_Stab_LyapFun}}}: 
The proof is done by showing that $V^{i}\big(.,v^i(.)\big)$ is a Lyapunov function for $v^{i}(.)$, for each given $i$. 
Denoting the value function of the initial admissible triggering policy with $V^0(.,.)$, it is continuous by definition of admissibility, and positive definite by positive semi-definiteness of $U(.,.)$ and Assumption \ref{Assum_InvariantSet}. Note that, there is no $x \neq 0$ with the value function of zero under any triggering policy. If $V^i(.,.)$ for some $i$ is positive definite, it directly follows from (\ref{eq_ADVI}) that $V^{i+1}(.,.)$ will also be positive definite, because, if $U(x,v)=0$ for some $x \neq 0$ and $v$, then $F_v(x) \neq x$ by Assumption \ref{Assum_InvariantSet}. Hence, by induction, $V^{i+1}(.,.)$ is positive definite for every $i \in \mathbb{N}$. Also, as shown in the proof of Lemma \ref{Lemma_Continuity_SADVI} it is a continuous function in $\Omega$. 
By (\ref{eq_ADVI})
\begin{equation}
\begin{split}
		V^{i+1}\big(&x,v^{i+1}(x)\big) = U\big(x,v^{i+1}(x)\big) + \\
		&V^i\Big(F_{v^{i+1}(x)}(x), v^i\big(F_{v^{i+1}(x)}(x)\big)\Big), \forall x \in \Omega. \label{Thm1_eq1}
\end{split}
\end{equation}
One has $V^{i+1}\big(x,v^{i+1}(x)\big) \leq V^{i+1}\big(x,v^{i}(x)\big)$, by (\ref{eq_ADVI_vi}). Moreover, $V^{i+1}\big(x,v^{i}(x)\big) \leq V^{i}\big(x,v^{i}(x)\big)$ by Lemma \ref{Lemma_NonDecreasing}. Hence,
\begin{equation}
		V^{i+1}\big(x,v^{i+1}(x)\big) \leq V^{i}\big(x,v^{i}(x)\big), \forall x \in \Omega, \forall i. \label{Thm1_eq1_2}
\end{equation}
Therefore, 
\begin{equation}
\begin{split}
		V^{i}\big(x,&v^{i}(x)\big) \geq U\big(x,v^{i+1}(x)\big) \\
		&+ V^i\Big(F_{v^{i+1}(x)}(x), v^i\big(F_{v^{i+1}(x)}(x)\big)\Big), \forall x \in \Omega, \label{Thm1_eq2}
\end{split}
\end{equation}
which leads to
\begin{equation}
\begin{split}
		\Delta V^i(x) &:= V^i\Big(F_{v^{i+1}(x)}(x), v^i\big(F_{v^{i+1}(x)}(x)\big)\Big) \\
		&- V^{i}\big(x,v^{i}(x)\big) \leq - U\big(x,v^{i+1}(x)\big), \forall x \in \Omega. \label{Thm1_eq3}
\end{split}
\end{equation}
Hence, the asymptotic stability of the system operated by $v^{i}(.)$ follows considering Assumption \ref{Assum_InvariantSet} and LaSalle's invariance theorem, \cite{Khalil}.

Set $\mathcal{B}^{i}_{\bar{r}^i}$ is an estimation of the region of attraction (EROA) \cite{Khalil} for the closed loop system, because, $V^{i}\big(x_{k+1}, v^i(x_{k+1})\big) \leq V^i\big(x_k, v^i(x_k\big)$ by (\ref{Thm1_eq3}), hence, $x_k\in\mathcal{B}^{i}_{\bar{r}^i}$ leads to $x_{k+1}\in\mathcal{B}^{i}_{\bar{r}^i}, \forall k \in \mathbb{N}$. 
Finally, since $\mathcal{B}^i_{\bar{r}^i}$ is contained in $\Omega$, it is bounded. Also, the set is closed, because, it is the \textit{inverse image} of a closed set, namely $[0,\bar{r}^i]$ under a continuous function, \cite{Rudin}. Hence, $\mathcal{B}^i_{\bar{r}^i}$ is compact. The origin is an \textit{interior} point of the EROA, because $V^i\big(0,v^i(0)\big)=0$, $\bar{r}^i>0$, and $V^i\big(.,v^i(.)\big)\in\mathcal{C}(\Omega)$.
\qed

\textit{\textbf{Proof of Theorem \ref{Thm_Stabil_SADVI_No_Lyap}}}: 
Eqs. (\ref{eq_ADVI}) and (\ref{eq_ADVI_vi}) and the monotonicity feature established in Lemma \ref{Lemma_NonDecreasing} lead to
\begin{equation}
\begin{split}
		V^{2}\big(x_0^*,v^0(x_0^*)\big) = &U\big(x_0^*,v^0(x_0^*)\big) + \\
		V^1\Big(F_{v^0(x_0^*)}&(x_0^*),v^1\big(F_{v^0(x_0^*)}(x_0^*)\big)\Big) \leq \\\
		&V^1\big(x_0^*,v^0(x_0^*)\big), \forall x_0^* \in \Omega, \label{Thm_Stab_NoLyap_1}
\end{split}
\end{equation}
and similarly
\begin{equation}
\begin{split}
		V^{3}\big(x_1^*,v^1(x_1^*)&\big) = U\big(x_1^*,v^1(x_1^*)\big) + \\
		V^2\Big(F_{v^1(x_1^*)}&(x_1^*),v^2\big(F_{v^1(x_1^*)}(x_1^*)\big)\Big) \leq \\
		V^2\big(x_1^*,&v^1(x_1^*)\big) \leq  V^1\big(x_1^*,v^1(x_1^*)\big), \forall x_1^* \in \Omega, \label{Thm_Stab_NoLyap_2}
\end{split}
\end{equation}
Let $x_{k+1}^*:=F_{v^{k}(x_{k}^*)}(x_{k}^*), \forall k \in \mathbb{N}$ and $x_0^*:=x_0$. Replacing $V^1\big(x_1^*,v^1(x_1^*)\big)$ in the left hand side of the inequality in (\ref{Thm_Stab_NoLyap_1}) with the left hand side of (\ref{Thm_Stab_NoLyap_2}), which is smaller per (\ref{Thm_Stab_NoLyap_2}), one has
\begin{equation}
\begin{split}
		U\big(x_0^*,v^0(x_0^*)\big) &+ U\big(x_1^*,v^1(x_1^*)\big) + V^2\big(x_2^*,v^2(x_2^*)\big) \leq \\
		& V^1\big(x_0^*,v^0(x_0^*)\big), \forall x_0^* \in \Omega, \forall v \in \{0,1\}, \label{Thm_Stab_NoLyap_3}
\end{split}
\end{equation}
Repeating this process by replacing $V^2\big(x_2^*,v^2(x_2^*)\big)$ in (\ref{Thm_Stab_NoLyap_3}) using
\begin{equation}
\begin{split}
		V^{4}\big(x_2^*,v^2(x_2^*)&\big) = U\big(x_2^*,v^2(x_2^*)\big) + \\
		V^3\Big(F_{v^2(x_2^*)}&(x_2^*),v^3\big(F_{v^2(x_2^*)}(x_2^*)\big)\Big) \leq \\
		V^3\big(x_2^*,&v^2(x_2^*)\big) \leq  V^2\big(x_2^*,v^2(x_2^*)\big), \forall x_2^* \in \Omega, \label{Thm_Stab_NoLyap_4}
\end{split}
\end{equation}
leads to
\begin{equation}
\begin{split}
		U\big(x_0^*,&v^0(x_0^*)\big) + U\big(x_1^*,v^1(x_1^*)\big) + U\big(x_2^*,v^2(x_2^*)\big) + \\ 
		&V^3\big(x_3^*,v^3(x_3^*)\big) \leq V^1\big(x_0^*,v^0(x_0^*)\big), \forall x_0^* \in \Omega, \forall v. \label{Thm_Stab_NoLyap_5}
\end{split}
\end{equation}
Similarly by repeating this process one has
\begin{equation}
\begin{split}
		\sum_{k=0}^{i-1} & U\big(x_k^*,v^{k}(x_k^*)\big) + V^{i}\big(x_i^*, v^{i}(x_i^*)\big) \\
		 & \leq V^1\big(x_0^*,v^0(x_0^*)\big), \forall x_0^* \in \Omega, \forall i \in \mathbb{N}-\{0\}. \label{Thm_Stab_NoLyap_6}
\end{split}
\end{equation}
Since $V^{i}(.,.)$ is positive definite, it can be dropped from the left hand side of the foregoing inequality. The result is, the sequence of partial sums in the left hand side is upper bounded by the right hand side and because of being non-decreasing, it converges, as $i\to\infty$, \cite{Rudin}. Therefore, $U\big(x_k^*,v^{k}(x_k^*)\big) \to 0$ as $k\to\infty$. Considering Assumption \ref{Assum_InvariantSet}, this leads to $x_k^* \to 0$, as long as the entire state trajectory is contained in $\Omega$. 
\qed

\textit{\textbf{Proof of Theorem \ref{Thm_ROA_Evolving_VI}}}: 
As the first step we show that for any given $r$ one has
\begin{equation}
x_k \in \mathcal{B}^{k}_{r} \Rightarrow x_{k+1}=F_{v^{k}(x_k)}(x_k) \in \mathcal{B}^{k+1}_{r}, \forall k \in \mathbb{N}, \forall r \in\mathbb{R}^+. \label{eq_Thm_ROA_Evolving_VI_1}
\end{equation}
From (\ref{Thm1_eq3}) one has $V^{k}\big(x_{k+1}, v^{k}(x_{k+1})\big) \leq V^{k}\big(x_{k}, v^{k}(x_{k})\big)$. Therefore, 
\begin{equation}
x_k \in \mathcal{B}^{k}_{r} \Rightarrow x_{k+1} \in \mathcal{B}^{k}_{r}, \forall k \in \mathbb{N}, \forall r \in\mathbb{R}^+. \label{eq_Thm_ROA_Evolving_VI_2}
\end{equation}
By (\ref{Thm1_eq1_2}) and the definition of $\mathcal{B}^{k}_r$ one has $\mathcal{B}^{k}_r \subset \mathcal{B}^{k+1}_r$. Therefore, 
\begin{equation}
x_{k+1} \in \mathcal{B}^{k}_{r} \Rightarrow x_{k+1} \in \mathcal{B}^{k+1}_{r}, \forall k \in \mathbb{N}, \forall r \in\mathbb{R}^+. \label{eq_Thm_ROA_Evolving_VI_3}
\end{equation}
Finally (\ref{eq_Thm_ROA_Evolving_VI_2}) and (\ref{eq_Thm_ROA_Evolving_VI_3}) lead to (\ref{eq_Thm_ROA_Evolving_VI_1}). Now that (\ref{eq_Thm_ROA_Evolving_VI_1}) is proved, one may use mathematical induction to see
\begin{equation}
x_0 \in \mathcal{B}^{0}_{r} \Rightarrow x_{k} \in \mathcal{B}^{k}_{r}, \forall k \in \mathbb{N}, \forall r \in\mathbb{R}^+. \label{eq_Thm_ROA_Evolving_VI_3_1}
\end{equation}
The next step is noting that $V^{*}\big(x,v^*(x)\big) \leq V^{k}\big(x,v^k(x)\big), \forall x$. This inequality leads to $\mathcal{B}^{k}_r \subset \mathcal{B}^{*}_r, \forall k$, by definition of $\mathcal{B}^{k}_r$ and $\mathcal{B}^{*}_r$. Therefore, (\ref{eq_Thm_ROA_Evolving_VI_3_1}) leads to 
\begin{equation}
x_0 \in \mathcal{B}^{0}_{r} \Rightarrow x_{k} \in \mathcal{B}^{*}_{r}, \forall k \in \mathbb{N}, \forall r \in\mathbb{R}^+. \label{eq_Thm_ROA_Evolving_VI_4}
\end{equation}
The result given by (\ref{eq_Thm_ROA_Evolving_VI_4}) proves the theorem, because, if $r$ is such that $\mathcal{B}^{*}_{r} \subset \Omega$ then any trajectory initiated within $\mathcal{B}^{0}_{r}$ will remain inside $\Omega$, and hence, by Theorem \ref{Thm_Stabil_SADVI_No_Lyap} will converge to the origin.
\qed

\vspace{10pt}

\begin{wrapfigure}{l}{0.15\textwidth}
  \vspace{-10pt}
  \begin{center}
		\includegraphics[width=0.3\columnwidth]{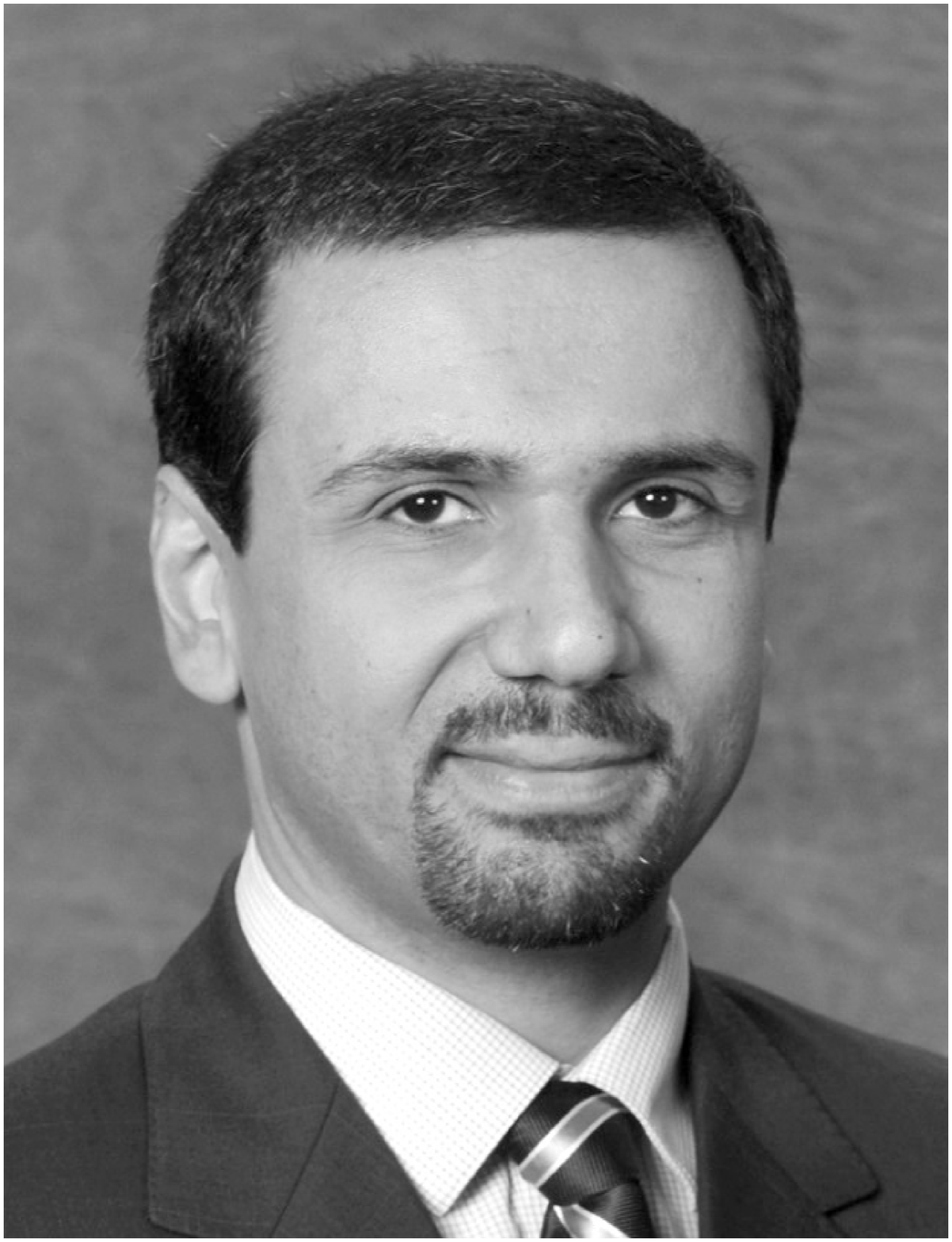}
  \end{center}
  \vspace{-0pt}
  \vspace{-10pt}
\end{wrapfigure}

\footnotesize
Ali Heydari received his PhD degree from the Missouri University of Science and Technology in 2013. He is currently an assistant professor of mechanical engineering at the South Dakota School of Mines and Technology. He was the recipient of the Outstanding M.Sc. Thesis Award from the Iranian Aerospace Society, the Best Student Paper Runner-Up Award from the AIAA Guidance, Navigation and Control Conference, and the Outstanding Graduate Teaching Award from the Academy of Mechanical and Aerospace Engineers at Missouri S\&T. His research interests include optimal control, approximate dynamic programming, and control of hybrid and switching systems. He is a member of Tau Beta Pi.


\begin{thebibliography}{10}

\bibitem{Hristu_LimitedCommunication}
D.~Hristu and K.~Morgansen, ``Limited communication control,'' {\em Systems \&
  Control Letters}, vol.~37, no.~4, pp.~193 -- 205, 1999.

\bibitem{Halevi_NSC_1988_DelayMainly}
Y.~Halevi and A.~Ray, ``Integrated communication and control systems. part 1 -
  analysis,'' {\em Journal of Dynamic Systems, Measurement and Control,
  Transactions of the ASME}, vol.~110, no.~4, pp.~367--373, 1988.

\bibitem{Walsh_NSC_ContSystMagazine}
G.~Walsh and H.~Ye, ``Scheduling of networked control systems,'' {\em IEEE
  Control Systems Magazine}, vol.~21, pp.~57--65, Feb 2001.

\bibitem{Yook_Event_Driven_StateTHreshold}
J.~Yook, D.~Tilbury, and N.~Soparkar, ``Trading computation for bandwidth:
  reducing communication in distributed control systems using state
  estimators,'' {\em IEEE Transactions on Control Systems Technology}, vol.~10,
  pp.~503--518, Jul 2002.

\bibitem{Ishii_SwitchBox}
H.~Ishii and B.~A. Francis, ``Stabilization with control networks,'' {\em
  Automatica}, vol.~38, no.~10, pp.~1745 -- 1751, 2002.

\bibitem{Antsaklis_PeriodicTriggering_ModelBased}
L.~Montestruque and P.~Antsaklis, ``State and output feedback control in
  model-based networked control systems,'' in {\em IEEE Conference on Decision
  and Control}, vol.~2, pp.~1620--1625, 2002.

\bibitem{Yang_NCS_ShortSurvey}
T.-C. Yang, ``Networked control system: a brief survey,'' {\em Control Theory
  and Applications, IEE Proceedings -}, vol.~153, pp.~403--412, July 2006.

\bibitem{Antsaklis_NCS_SpecialIssue}
P.~Antsaklis and J.~Baillieul, ``Special issue on technology of networked
  control systems,'' {\em Proceedings of the IEEE}, vol.~95, pp.~5--8, Jan
  2007.

\bibitem{Gupta_NCS_Overview}
R.~Gupta and M.-Y. Chow, ``Networked control system: Overview and research
  trends,'' {\em IEEE Transactions on Industrial Electronics}, vol.~57,
  pp.~2527--2535, July 2010.

\bibitem{Antsaklis_MB_NCS_EVentTriggering}
E.~Garcia and P.~Antsaklis, ``Model-based event-triggered control for systems
  with quantization and time-varying network delays,'' {\em IEEE Transactions
  on Automatic Control}, vol.~58, pp.~422--434, Feb 2013.

\bibitem{NCS_SmartGrid}
H.~Li, J.~B. Song, and Q.~Zeng, ``Adaptive modulation in networked control
  systems with application in smart grids,'' {\em IEEE Communications Letters},
  vol.~17, pp.~1305--1308, July 2013.

\bibitem{WirelessSensorNetwork_Airplane}
R.~Yedavalli and R.~Belapurkar, ``Application of wireless sensor networks to
  aircraft control and health management systems,'' {\em Journal of Control
  Theory and Applications}, vol.~9, no.~1, pp.~28--33, 2011.

\bibitem{Nesic_QCS_NCS}
D.~Nesic and D.~Liberzon, ``A unified framework for design and analysis of
  networked and quantized control systems,'' {\em IEEE Transactions on
  Automatic Control}, vol.~54, pp.~732--747, April 2009.

\bibitem{Astrom_ProposingEventTriggering}
K.~Astrom and B.~Bernhardsson, ``Comparison of {Riemann} and {Lebesgue}
  sampling for first order stochastic systems,'' in {\em Proceedings of the
  IEEE Conference on Decision and Control}, vol.~2, pp.~2011--2016, Dec 2002.

\bibitem{Walsh_TOD}
G.~Walsh, H.~Ye, and L.~Bushnell, ``Stability analysis of networked control
  systems,'' {\em IEEE Transactions on Control Systems Technology}, vol.~10,
  pp.~438--446, May 2002.

\bibitem{Tabuada_EventTriggering}
P.~Tabuada, ``Event-triggered real-time scheduling of stabilizing control
  tasks,'' {\em IEEE Transactions on Automatic Control}, vol.~52,
  pp.~1680--1685, Sept 2007.

\bibitem{Astorm_GZOH}
K.~J. Astrom, ``Event based control,'' in {\em Analysis and Design of Nonlinear
  Control Systems} (A.~Astolfi and L.~Marconi, eds.), Springer Berlin
  Heidelberg, 2008.

\bibitem{Rabi_OptimalControl_NCS}
M.~Rabi, K.~Johansson, and M.~Johansson, ``Optimal stopping for event-triggered
  sensing and actuation,'' in {\em IEEE Conference on Decision and Control},
  pp.~3607--3612, Dec 2008.

\bibitem{Heemels_IJC_ZOH}
W.~P. M.~H. Heemels, J.~H. Sandee, and P.~P.~J. Van Den~Bosch, ``Analysis of
  event-driven controllers for linear systems,'' {\em International Journal of
  Control}, vol.~81, no.~4, pp.~571--590, 2008.

\bibitem{Lunze_PerfectModel_GZOH}
J.~Lunze and D.~Lehmann, ``A state-feedback approach to event-based control,''
  {\em Automatica}, vol.~46, no.~1, pp.~211--215, 2010.

\bibitem{Wang_ISS}
X.~Wang and M.~Lemmon, ``On event design in event-triggered feedback systems,''
  {\em Automatica}, vol.~47, no.~10, pp.~2319 -- 2322, 2011.

\bibitem{Marchand_ZOH}
N.~Marchand, S.~Durand, and J.~Castellanos, ``A general formula for event-based
  stabilization of nonlinear systems,'' {\em IEEE Transactions on Automatic
  Control}, vol.~58, pp.~1332--1337, May 2013.

\bibitem{Cassandras_CCC}
C.~Cassandras, ``Event-driven control, communication, and optimization,'' in
  {\em Chinese Control Conference, CCC}, pp.~1--5, 2013.

\bibitem{Sahoo_Jagannathan_ACC13}
A.~Sahoo, H.~Xu, and S.~Jagannathan, ``Neural network-based adaptive
  event-triggered control of affine nonlinear discrete time systems with
  unknown internal dynamics,'' in {\em American Control Conference}, 2013.

\bibitem{Tatikonda_EncoderDecoder}
S.~Tatikonda and S.~Mitter, ``Control under communication constraints,'' {\em
  IEEE Transactions on Automatic Control}, vol.~49, pp.~1056--1068, 2004.

\bibitem{Premaratne_EventTriggering_SpeechCoding}
U.~Premaratne, S.~Halgamuge, and I.~Mareels, ``Event triggered adaptive
  differential modulation: A new method for traffic reduction in networked
  control systems,'' {\em IEEE Transactions on Automatic Control}, vol.~58,
  pp.~1696--1706, July 2013.

\bibitem{Goktas_Delay_RobustControl}
F.~Goktas, J.~Smith, and R.~Bajcsy, ``$\mu$-synthesis for distributed control
  systems with network-induced delays,'' in {\em IEEE Conference on Decision
  and Control}, vol.~1, pp.~813--814, 1996.

\bibitem{Nilsson_LQG_forDelays_TimeStamping}
J.~Nilsson, B.~Bernhardsson, and B.~Wittenmark, ``Stochastic analysis and
  control of real-time systems with random time delays,'' {\em Automatica},
  vol.~34, no.~1, pp.~57 -- 64, 1998.

\bibitem{Wu_NCS_Delay}
J.~Wu, F.-Q. Deng, and J.-G. Gao, ``Modeling and stability of long random delay
  networked control systems,'' in {\em International Conference on Machine
  Learning and Cybernetics}, vol.~2, pp.~947--952, Aug 2005.

\bibitem{Yue_RobustCOntrol_For_Delayand_Droppout}
D.~Yue, Q.-L. Han, and J.~Lam, ``Network-based robust ${H}_{\infty}$ control of
  systems with uncertainty,'' {\em Automatica}, vol.~41, pp.~999 -- 1007, 2005.

\bibitem{Kim_MaximumAllowableDelayBound}
D.-S. Kim, Y.~S. Lee, W.~H. Kwon, and H.~S. Park, ``Maximum allowable delay
  bounds of networked control systems,'' {\em Control Engineering Practice},
  vol.~11, no.~11, pp.~1301 -- 1313, 2003.

\bibitem{Yi_BackProp_ForDelayApprox}
J.~Yi, Q.~Wang, D.~Zhao, and J.~T. Wen, ``{BP} neural network prediction-based
  variable-period sampling approach for networked control systems,'' {\em
  Applied Mathematics and Computation}, vol.~185, pp.~976 -- 988, 2007.

\bibitem{Du_AMC_AdaptiveCritic_StochasticSystem}
D.~Du and M.~Fei, ``A two-layer networked learning control system using
  actor–critic neural network,'' {\em Applied Mathematics and Computation},
  vol.~205, no.~1, pp.~26 -- 36, 2008.

\bibitem{Xu_Jagannathan_Automatica}
H.~Xu, S.~Jagannathan, and F.~Lewis, ``Stochastic optimal control of unknown
  linear networked control system in the presence of random delays and packet
  losses,'' {\em Automatica}, vol.~48, pp.~1017 -- 1030, 2012.

\bibitem{Repele_Delay_SmithPredictor_DelayCalculation}
L.~Repele, R.~Muradore, D.~Quaglia, and P.~Fiorini, ``Improving performance of
  networked control systems by using adaptive buffering,'' {\em IEEE
  Transactions on Industrial Electronics}, vol.~61, pp.~4847--4856, 2014.

\bibitem{Delchamps_Quantization}
D.~F. Delchamps, ``Stabilizing a linear system with quantized state feedback,''
  {\em IEEE Transactions on Automatic Control}, vol.~35, pp.~916--924, Aug
  1990.

\bibitem{Brockett_Quantization}
R.~Brockett and D.~Liberzon, ``Quantized feedback stabilization of linear
  systems,'' {\em IEEE Transactions on Automatic Control}, vol.~45,
  pp.~1279--1289, Jul 2000.

\bibitem{Antsaklis_PeriodicTriggering_With_TimeVaryingPeriod}
L.~Montestruque and P.~Antsaklis, ``Stability of model-based networked control
  systems with time-varying transmission times,'' {\em IEEE Transactions on
  Automatic Control}, vol.~49, pp.~1562--1572, Sept 2004.

\bibitem{Xu_CDC2004_OptimalCommunicationPolicy}
Y.~Xu and J.~Hespanha, ``Optimal communication logics in networked control
  systems,'' vol.~4, pp.~3527--3532, 2004.

\bibitem{Farokhi_TAC}
F.~Farokhi and K.~Johansson, ``Stochastic sensor scheduling for networked
  control systems,'' {\em IEEE Transactions on Automatic Control}, vol.~59,
  pp.~1147--1162, May 2014.

\bibitem{Pappas_TAC_OptimalPOwerManagement}
K.~Gatsis, A.~Ribeiro, and G.~Pappas, ``Optimal power management in wireless
  control systems,'' {\em IEEE Transactions on Automatic Control}, vol.~59,
  pp.~1495--1510, June 2014.

\bibitem{Werbos_ActionDependent_Critic}
P.~Werbos, ``Neural networks for control and system identification,'' in {\em
  Proceedings of the 28th IEEE Conference on Decision and Control},
  pp.~260--265, 1989.

\bibitem{Bertsekas_NDP}
D.~P. Bertsekas and J.~N. Tsitsiklis, {\em Neuro-Dynamic Programming}.
\newblock Athena Scientific, 1996.

\bibitem{Watkins}
C.~Watkins, {\em Learning from Delayed Rewards}.
\newblock PhD Dissertation, Cambridge University, Cambridge, England, 1989.

\bibitem{Sutton}
R.~S. Sutton and A.~G. Barto, {\em Reinforcement Learning: An Introduction}.
\newblock MIT Press, 1998.

\bibitem{Heydari_Franklin}
A.~Heydari and S.~Balakrishnan, ``Optimal switching between autonomous
  subsystems,'' {\em Journal of the Franklin Institute}, vol.~351,
  pp.~2675--2690, 2014.

\bibitem{HaiboHe_NCS}
X.~Zhong, Z.~Ni, H.~He, X.~Xu, and D.~Zhao, ``Event-triggered reinforcement
  learning approach for unknown nonlinear continuous-time system,'' in {\em
  Int. Joint Conf. on Neural Networks}, pp.~3677--3684, 2014.

\bibitem{Ferrari_EventTriggering}
D.~Tolic, R.~Fierro, and S.~Ferrari, ``Optimal self-triggering for nonlinear
  systems via approximate dynamic programming,'' in {\em IEEE Int. Conf. on
  Control Applications}, pp.~879--884, 2012.

\bibitem{Kirk}
D.~E. Kirk, {\em Optimal control theory; an introduction}.
\newblock Prentice-Hall, 1998.

\bibitem{Weierstrass_Theorem}
H.~Jeffreys and B.~S. Jeffreys, ``Weierstrass's theorem on approximation by
  polynomials,'' in {\em Methods of Mathematical Physics}, pp.~446--448,
  Cambridge University Press, 3rd~ed., 1988.

\bibitem{Hornik_NN_Continuity}
K.~Hornik, M.~Stinchcombe, and H.~White, ``Multilayer feedforward networks are
  universal approximators,'' {\em Neural Networks}, vol.~2, no.~5,
  pp.~359--366, 1989.

\bibitem{TCP_Book}
W.~R. Stevens, {\em TCP/IP Illustrated, Vol. 1: The Protocols}.
\newblock Addison-Wesley Professional, 1st~ed., 1993.

\bibitem{Howard_PI_MDP}
R.~Howard, {\em Dynamic Programming and Markov Processes}.
\newblock MIT Press, Cambridge, MA, 1960.

\bibitem{Gosavi_StocasticDP}
A.~Gosavi, ``Control optimization with stochastic dynamic programming,'' in
  {\em Simulation-Based Optimization}, pp.~133--210, Springer, 2003.

\bibitem{Werbos2012}
P.~J. Werbos, ``Reinforcement learning and approximate dynamic programming
  {(RLADP)}-foundations, common misconceptions, and the challenges ahead,'' in
  {\em Reinforcement Learning and Approximate Dynamic Programming for Feedback
  Control} (F.~L. Lewis and D.~Liu, eds.), pp.~1--30, John Wiley \& Sons, 2012.

\bibitem{Bertsekas2012}
D.~P. Bertsekas, ``Lambda-policy iteration: A review and a new
  implementation,'' in {\em Reinforcement Learning and Approximate Dynamic
  Programming for Feedback Control} (F.~L. Lewis and D.~Liu, eds.),
  pp.~381--406, John Wiley \& Sons, 2012.

\bibitem{AlTamimi}
A.~Al-Tamimi, F.~Lewis, and M.~Abu-Khalaf, ``Discrete-time nonlinear {HJB}
  solution using approximate dynamic programming: Convergence proof,'' {\em
  IEEE Trans. Systems, Man, and Cybernetics, Part B: Cybernetics}, vol.~38,
  pp.~943--949, Aug 2008.

\bibitem{Rudin}
W.~Rudin, {\em Principles of Mathematical Analysis}.
\newblock McGraw-Hill, 3rd~ed., 1976.
\newblock pp. 55, 60, 89.

\bibitem{Heydari_TCYB}
A.~Heydari, ``Revisiting approximate dynamic programming and its convergence,''
  {\em IEEE Transactions on Cybernetics}, vol.~44, pp.~2733--2743, 2014.

\bibitem{Heydari_Franklin2}
A.~Heydari, ``Optimal scheduling for reference tracking or state regulation
  using reinforcement learning,'' {\em Journal of the Franklin Institute}.
\newblock in press and available online.

\bibitem{Lincol_RelaxingDynProg}
B.~Lincoln and A.~Rantzer, ``Relaxing dynamic programming,'' {\em IEEE
  Transactions on Automatic Control}, vol.~51, pp.~1249--1260, Aug 2006.

\bibitem{Rinehart_VI_TAC}
M.~Rinehart, M.~Dahleh, and I.~Kolmanovsky, ``Value iteration for (switched)
  homogeneous systems,'' {\em IEEE Transactions on Automatic Control}, vol.~54,
  no.~6, pp.~1290--1294, 2009.

\bibitem{Khalil}
H.~Khalil, {\em Nonlinear Systems}.
\newblock Prentice-Hall, 2002.
\newblock pp. 111-181.

\bibitem{Heydari_SAVI}
A.~Heydari, ``Stabilizing value iteration with and without approximation
  errors,'' available at arXiv.org.

\end{thebibliography}
\end{document}